\documentclass[journal]{IEEEtran}
\usepackage{cite}
\usepackage{amsmath,amssymb}
\usepackage{multicol}
\usepackage{graphicx}
\usepackage{amsmath}
\usepackage{caption}
\usepackage{algorithm}
\usepackage{algpseudocode} % For algorithmic syntax
\usepackage{placeins}
\usepackage{comment}
\usepackage{float}
\usepackage{afterpage}
\usepackage{stfloats}
\usepackage{color}

\usepackage{xcolor}
\title{An Efficient Correlation-based Evaluation of the GN-Model for General Multi-Span Optical Systems}

\author{
        Y.~Jiang, Y.~Gao, P.~Poggiolini, \IEEEmembership{Fellow, IEEE}
\thanks{Y. Jiang and P. Poggiolini are with the Dipartimento di Elettronica e Telecomunicazioni (DET), Politecnico di Torino, 10129, Torino, Italy. Y. Gao is with the Dipartimento di Elettronica e Telecomunicazioni (DET), Politecnico di Torino, 10129, Torino, Italy, and the school of electronic and information engineering, Soochow University, Suzhou, China. The three authors have contributed in equal measure to all aspects of the paper. All the key ideas were conceived by the authors and all the main calculations were carried out originally by the authors. AI was used to polish the text, verify the calculations and for limited formula extensions.} 

\thanks{This work was partially supported by Cisco Systems through the ``IDIMOS'' Sponsored Research Agreement (SRA) and by the PhotoNext Center of Politecnico di Torino. }}

\begin{document}

\maketitle
\begin{abstract}
Accurate and computationally efficient evaluation of non-linear interference (NLI) remains challenging in multi-band WDM systems comprising heterogeneous spans and nonidentical channels. In such systems, channels may have distinct spatial power profiles and multi-subcarrier structures. Spans may use different fiber types with span-dependent dispersion and loss profiles. These systems may further employ general amplification schemes, including forward and backward Raman amplification. Inter-channel stimulated Raman scattering and multiple lumped gain or loss elements may also be present.

In this encompassing framework, we provide an efficient decomposition of the GN-model NLI contributions. We show that the NLI power spectral density at any frequency across the WDM comb is fundamentally driven by the spatial auto- and cross-correlations of the channels' spatial power profiles or, equivalently, by the corresponding energy spectra. For this reason we call this formalism the `correlation GN-model', or cGN. Once these correlations or spectra have been computed, the remaining calculations reduces to a fast, well-behaved one-dimensional numerical integral. 
Moreover, cGN fully accounts for coherent interference of NLI in every pair of spans.  Such coherence is often neglected or approximated in existing analytical and semi-analytical GN-model formulations, although it can increase the NLI by several dB.

In short, the cGN framework applies to any general systems over any band and any bandwidth, with very high computational efficiency. It is capable of providing the frequency spectrum of NLI at multiple frequencies across each channel or subcarrier. Its accuracy is very high because virtually no approximations are invoked.

The cGN framework is available as freely downloadable software from the European Union Zenodo website.
 
\end{abstract}
    
\begin{IEEEkeywords}
GN-model, NLI PSD, coherence
\end{IEEEkeywords}
  
\section{Introduction}
\label{sec:Introduction}

Accurate modeling of nonlinear interference (NLI) is essential for the design and optimization of modern optical fiber systems\cite{2024_ECOC_Puttnam}--\cite{2025_OFC_Sarkis}, especially in challenging scenarios such as ultra-wideband, low-dispersion, and multi-subcarrier systems. 
The Gaussian Noise (GN) model \cite{2012_JLT_Poggiolini}--\cite{2022_JOCN_GNPy} and the Enhanced Gaussian Noise (EGN) model typically provide an adequately reliable reference for estimating NLI. However, their brute-force numerical evaluation is unsuitable for all those applications that require  computational efficiency and fast results. This limitation has motivated the development of closed-form and semi-analytical models that enable faster and more scalable NLI estimation \cite{2018_arXiv_Poggiolini} - \cite{2026_JLT_Poggiolini}. 

The computational efficiency of closed-form models (CFMs) is generally obtained through assumptions tailored to particular system conditions. Depending on the model, these may include, among others: i) evaluating the NLI power spectral density (PSD) only at the center frequency of the channel under test (CUT) and then assuming the PSD to be spectrally flat; ii) neglecting multi-channel interference (MCI); iii)  approximating the GN-model integration ``islands", which segment its two-dimensional frequency integration domain, with rectangles or stripes; iv) assuming incoherent accumulation of NLI across spans, or crudely approximating its coherence. In addition, Raman amplification, inter-channel stimulated Raman scattering (ISRS) and lumped loss are often not supported or approximated. 

All these approximations may be acceptable in conventional high-dispersion and high-symbol-rate systems, especially under simplifying assumptions such as all identical spans and all identical equally-spaced channels. But their accuracy may deteriorate in scenarios of practical interest, where such special conditions do not exist. Also, low-dispersion spans and/or multi-subcarrier channels systems are particularly challenging scenarios for CFMs.

In this work, we start out from the GN-model reference formula (GNRF) \cite{2014_JLT_Poggiolini}. We then use three key strategies to achieve efficient and accurate NLI evaluation. First, we develop a correlation-based formalism that reformulates part of the integrals into an analytically-interpretable form. Second, to preserve accuracy when the frequency integration plane is fractured into many oddly shaped islands, we develop an island-covering-strategy with rectangular subdomains whereby the error can be made arbitrarily small. This construction provides a common treatment of single-channel interference (SCI), cross-channel interference (XCI), and MCI. It also allows evaluating the NLI at arbitrary frequencies rather than only at the CUT center frequency for channels with arbitrary symbol rates, including multi-subcarrier channels.

%the signal PSDs and other frequency-dependent parameters are represented by local values, so that the accuracy of the overall calculation can be controlled through the rectangular partition.  

Third, we introduce an accumulated-dispersion representation that explicitly captures the phase evolution along the link. This leads to a natural decomposition of each NLI contribution into single-span (SS) and cross-span (XS) terms, both described by a common analytical structure. We show that the SS contributions are governed by the autocorrelation of the involved channel spatial power profiles (SPPs)  or, equivalently, by the corresponding energy spectrum. The XS contributions are described by cross-correlations and cross-spectra of each pair of spans, together with their relative phase-shift due to accumulated dispersion. This representation turns out to support span-dependent fiber parameters and general SPPs resulting from fiber attenuation, inter-channel stimulated Raman scattering, distributed Raman amplification, and lumped gain or loss. 

The GN model requires four nested integrals, of which two are space and two are frequency integrals. To deal with them, we have used two equivalent evaluation routes. In the first, the two space integrations are formally performed before the two frequency integrations, which are in turn reduced to a single-dimensional integral over a frequency-product variable. In the second, the  two frequency integrals are evaluated analytically through a coherence-response function, after which only a one-dimensional spatial-correlation integral remains. 
%For identical spans, the sum of the XScontributions can be further simplified through an analytical coherent-accumulation factor.

The proposed correlation GN-model (cGN) formulation is validated against direct numerical evaluation of the whole-link GN-model integral, hereafter referred to as the GN-model benchmark. The comparisons include single- and multi-channel configurations, frequency-resolved NLI spectra, low- and conventional-dispersion regimes, nonuniform channel arrangements, complex SPPs due to Raman amplification and ISRS, and coherent NLI accumulation over multi-span multi-band systems. The results show that the proposed cGN formulation reproduces the GN-model benchmark with high accuracy and high computational efficiency. At an accuracy of better than 0.1 dB  of NLI power estimation versus the GN-model benchmark, the cGN formulation is about 2000x faster.

The semi-analytical formalism proposed here can also be cast into a fully analytical solution by means of a polynomial expansion of the channels SPPs. This possibility, which borrows from the approach \cite{2026_JLT_Poggiolini}, is outside of the scope of this paper and will be dealt with in a separate submission.

The paper is organized as follows. Section~\ref{sect:NLI_accumulation} introduces the whole-link GN-model formulation, the rectangular decomposition of the frequency-integration domain, and the separation of the coherently accumulated NLI into SS and XS contributions. Section~\ref{sect:semi_analytical_SS} develops the two correlation-domain evaluation methods for the SS contributions. Section~\ref{sect:semi_analytical_XS} extends the formulation to XS contributions. 
%and discusses some simplifications obtained for identical spans. 
Section~\ref{sect:validation} validates the proposed formulation against the GN-model benchmark. Finally, the principal conclusions are summarized.

\begin{figure*}
    \centering
    \includegraphics[width=0.8\textwidth]{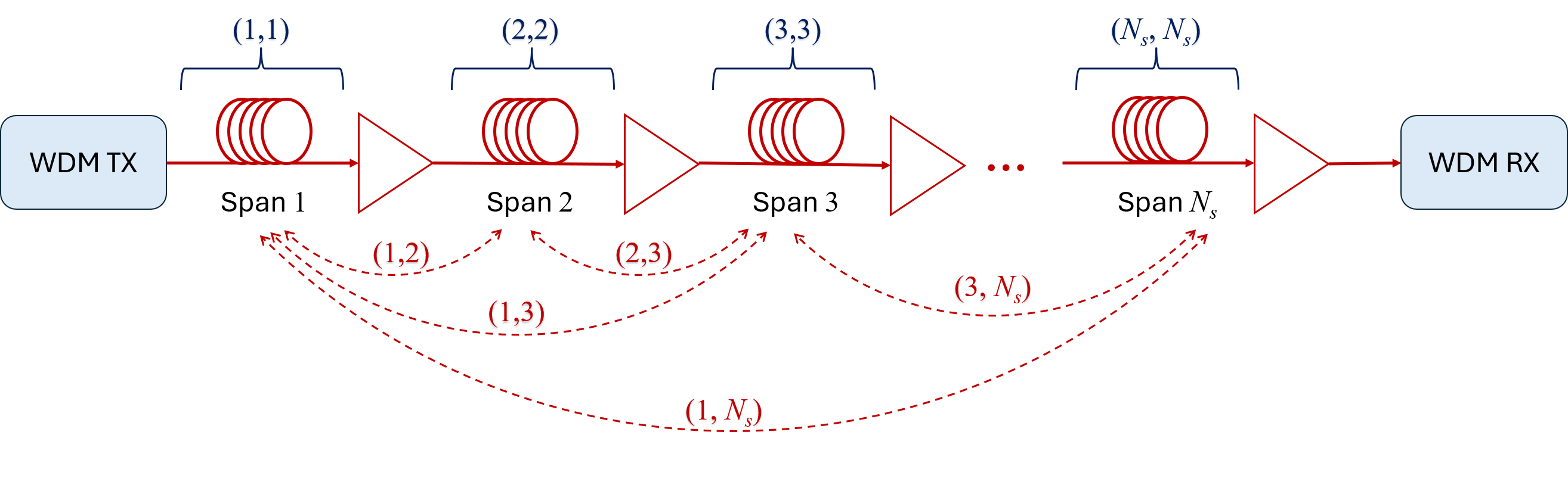}
    \caption{Illustration of coherent NLI accumulation in a multi-span link. The blue brackets denote SS contributions from any span $i$. The red dashed arrows denote XS contributions from the spans  $(i,j), i \neq j$. }
    \label{fig:coherent_NLI}
\end{figure*}

\section{GN Model and Span-Pair Decomposition}
\label{sect:NLI_accumulation}
We begin with the GNRF \cite{2014_JLT_Poggiolini}, which provides the NLI PSD at any frequency $f$ for dual-polarization transmission: 
\begin{equation}
    \begin{aligned}
        G&_{\mathrm{NLI}}(f)  =  \frac{16}{27} \int_{-\infty}^{\infty}  \int_{-\infty}^{\infty}
            G_{\mathrm{WDM}}(f_1) G_{\mathrm{WDM}}(f_2) \\
            & \times G_{\mathrm{WDM}}(f_1+f_2-f) \left| \rho_{\mathrm{FWM}}(f_1,f_2,f) \right|^2 df_1 df_2 
    \end{aligned}
    \label{eq:GNLI_GN}
\end{equation}
where $G_{\mathrm{WDM}}(f_i)$ denotes the WDM signal PSD, and $\rho_{\mathrm{FWM}}(f_1,f_2,f)$ denotes the Four-Wave Mixing (FWM) efficiency accumulated coherently from the transmitter to the receiver. The two elements that primarily determine the computational complexity are the evaluation of  $\rho_{\mathrm{FWM}}(f_1,f_2,f)$ and the geometry of frequency-integration domain in the $f_1-f_2$ plane.

\subsection{Integration domains and FWM efficiency}
\label{subsect:FWM_efficiency}
\begin{figure}
    \centering
    \includegraphics[width=0.33\textwidth]{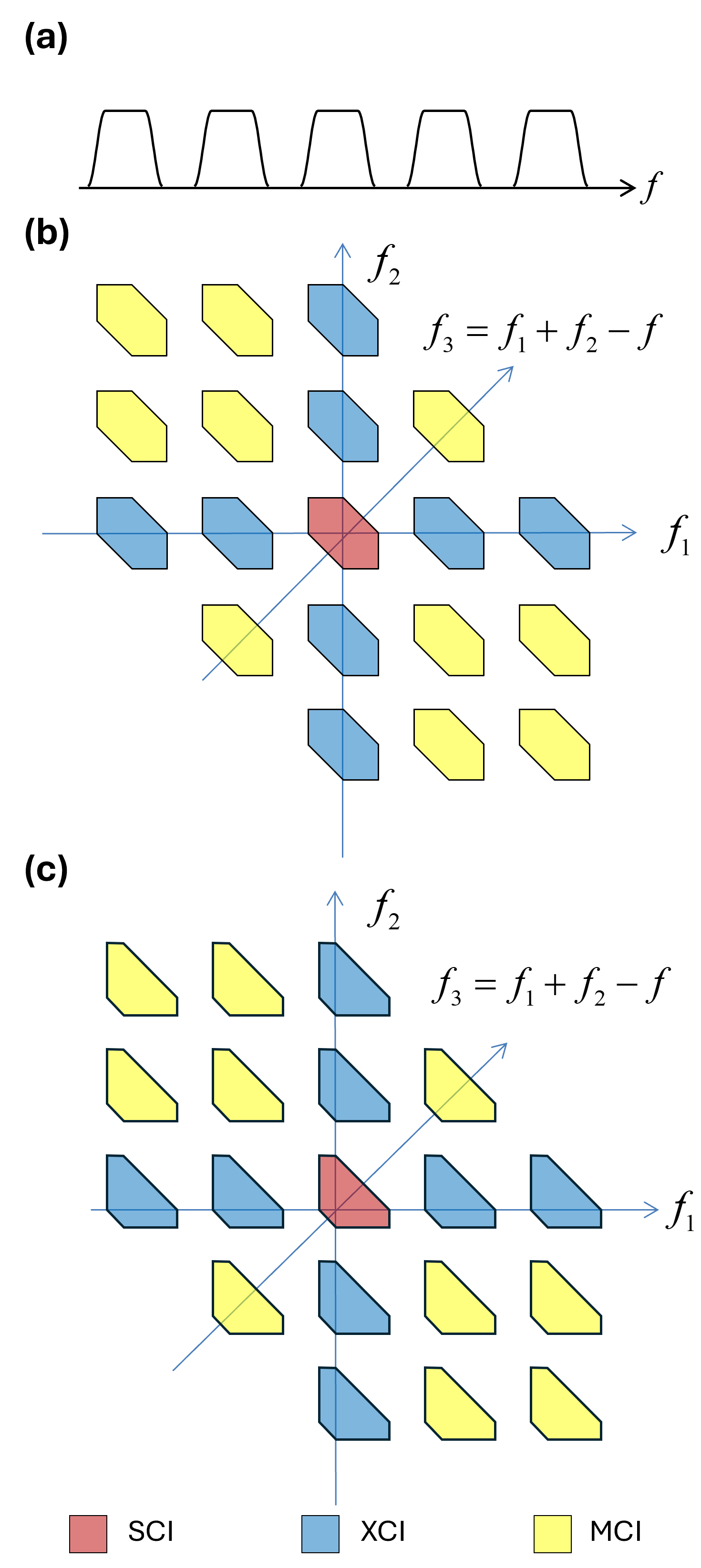}
    \caption{(a) a WDM signal PSD with 5 identical and equally spaced channels; GN-model integration islands for NLI PSD calculation (b) at the center frequency and (c) at a frequency to the left of the center frequency of the center channel.}
    \label{fig:islands_identical_5channels}
\end{figure}

The nonzero support of the three WDM signal PSDs in Eq.~(\ref{eq:GNLI_GN}) defines a set of bounded regions, commonly referred to as \textit{integration islands}, in the $(f_1,f_2)$ plane. For identical, equally spaced WDM channels, these islands are regular, as illustrated in Fig.~\ref{fig:islands_identical_5channels}. Fig.~\ref{fig:islands_identical_5channels}~(a) shows the WDM signal PSD, whereas Fig.~\ref{fig:islands_identical_5channels}~(b) and (c)  show the integration islands when the NLI PSD is evaluated at the center frequency and at a frequency to the left of the center, respectively. This comparison shows that the island geometry depends on the observation frequency. 
The FWM efficiency $\rho_{\mathrm{FWM}}(f_1,f_2,f)$ in Eq.~(\ref{eq:GNLI_GN}) describes the efficiency with which the interacting signal frequency components generate nonlinear distortion at frequency $f$. It depends on SPP, chromatic dispersion, fiber nonlinearity, span lengths, lumped gain/loss along the link. 

\begin{figure*}
    \centering
    \includegraphics[width=0.75\textwidth]{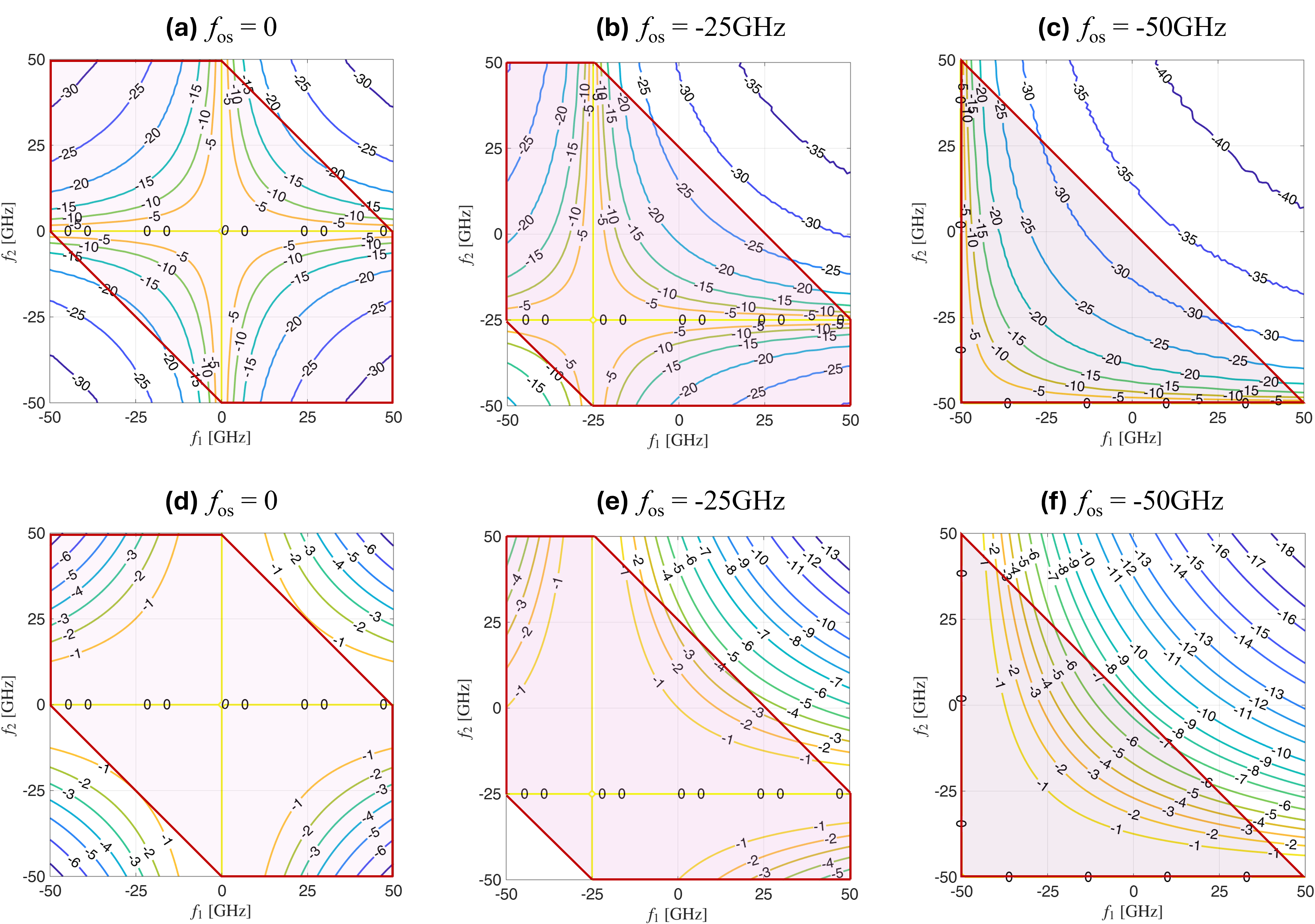}
    \caption{Normalized FWM efficiency (in dB scale) for SCI in a 100-km single-channel link with attenuation $\alpha_T=0.20~\mathrm{dB/km}$. Results are shown at frequency offsets (a),(d) $f_{\mathrm{os}}=0$, (b),(e) $f_{\mathrm{os}}=-25~\mathrm{GHz}$, and (c),(f) $f_{\mathrm{os}}=-50~\mathrm{GHz}$. The dispersion is $\beta_2=-20~\mathrm{ps^2/km}$ in (a)--(c) and $\beta_2=-1~\mathrm{ps^2/km}$ in (d)--(f). The shaded regions indicate the corresponding GN-model integration domains.}
    \label{fig:FWM_eff}
\end{figure*}

As illustrated in Fig.~\ref{fig:islands_identical_5channels}~(b) and (c), when the observation frequency moves aways from the channel center, the integration domains are progressively shifted and clipped, producing increasingly asymmetric shapes that may become triangular at a channel edge. Fig.~\ref{fig:FWM_eff} further illustrate the integration island for the SCI contribution at three different frequency offsets, together with the corresponding normalized FWM efficiency. Both the integration domain and the FWM efficiency depend on the observation frequency. Consequently, approximating the NLI PSD as constant across the channel can be inaccurate, particularly in low-dispersion spans and/or multi-subcarrier channels systems. The NLI PSD at $f_{\mathrm{os}}=-50~\mathrm{GHz}$ is approximately $5$--$6~\mathrm{dB}$ below its value at the channel center. This reduction results from both the modified integration domain and the frequency-dependent FWM efficiency. Here, $f_{\mathrm{os}}$ denotes the offset from the channel center frequency. Accordingly, the formulation developed below evaluates the NLI PSD at arbitrary frequencies and retains the frequency dependence of the GN-model integration domain.

\begin{figure}
    \centering
    \includegraphics[width=0.35\textwidth]{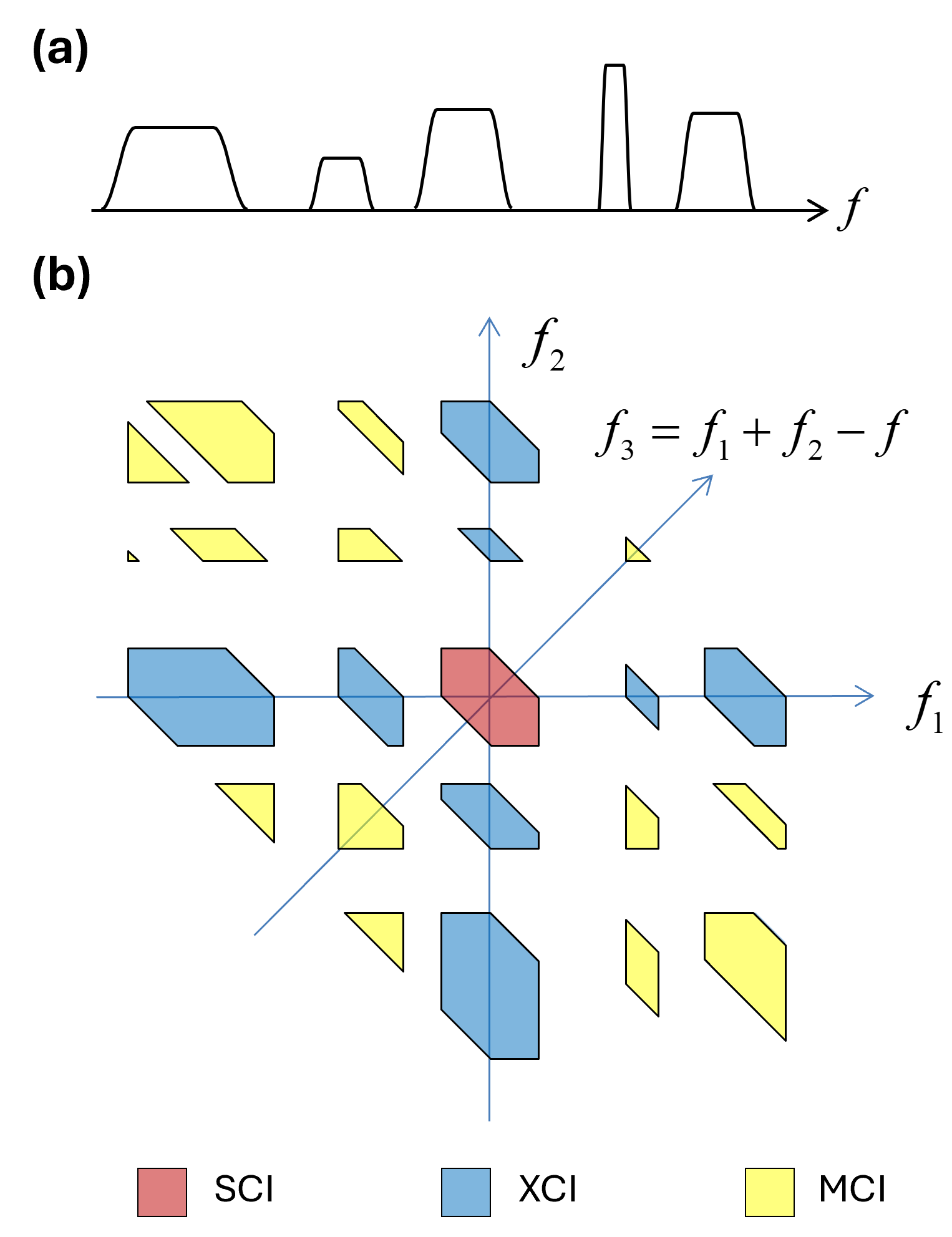}
    \caption{(a) a WDM signal PSD with 5 different channels; (b) GN-model integration islands for NLI PSD calculation at the center frequency of the center channel.}
    \label{fig:islands_diff_5channels}
\end{figure}

\begin{figure}
    \centering
    \includegraphics[width=0.8\columnwidth]{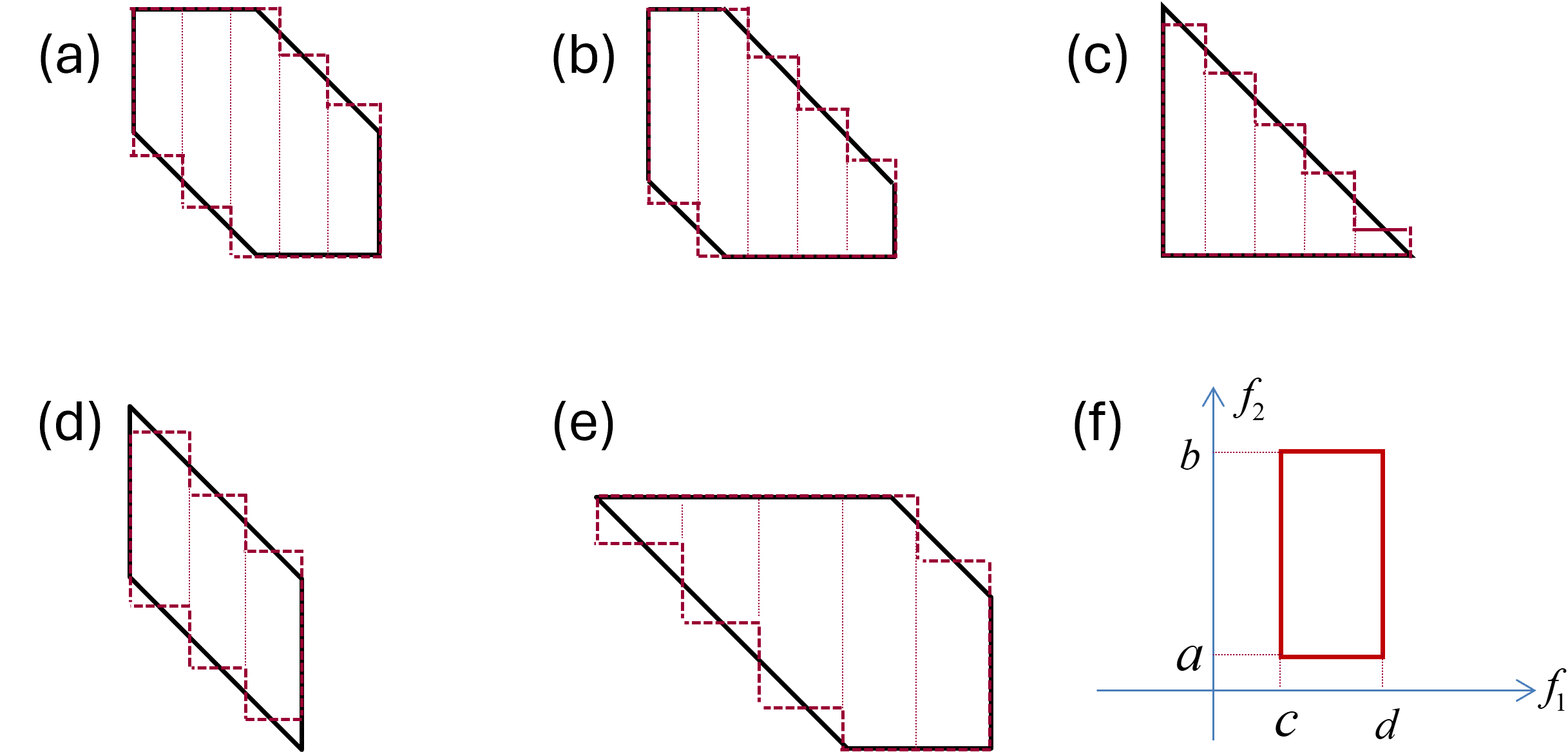}
    \caption{Black lines: Some typical GN-model integration islands. Red dashed lines: rectangles used to cover the islands for (a) any lozenge; (b), (c) other islands in systems similar to that in Fig.~\ref{fig:islands_identical_5channels}; and/or (d), (e) islands in system similar to that in Fig.~\ref{fig:islands_diff_5channels} ; (f) an arbitrary rectangle which can be placed anywhere in the plane.   }
    \label{fig:arbitrary_rectangle}
\end{figure}

For systems containing non-identical channels, such as channels with different symbol rates and unequal channel spacings, the integration islands become irregular and asymmetric, as shown in Fig.~\ref{fig:islands_diff_5channels}. These bounded domains can be approximated using a collection of non-overlapping rectangles to any desired accuracy, as shown in Fig.~\ref{fig:arbitrary_rectangle}~(a)-(e). Therefore, the fundamental problem is to solve the GN-model integral over a rectangle $[a,b]\times[c,d]$, placed anywhere in the $f_1-f_2$ plane, as shown in Fig.~\ref{fig:arbitrary_rectangle}~(f). The NLI contribution from one such rectangle is expressed as:
\begin{equation}
    \begin{aligned}
        G&_{\mathrm{NLI,rect}}(f)  =  \frac{16}{27} \int_{a}^{b}  \int_{c}^{d}
            G_{\mathrm{WDM}}(f_1) G_{\mathrm{WDM}}(f_2) \\
            & \times G_{\mathrm{WDM}}(f_1+f_2-f) \left| \rho_{\mathrm{FWM}}(f_1,f_2,f) \right|^2 {d}f_1 {d}f_2 
    \end{aligned}
    \label{eq:GNLI_GN_rect}
\end{equation}

Following the procedure used to obtain Eqs.~(10)-(12) in \cite{2026_JLT_Poggiolini}, and defining the shifted coordinates $f_1'=f_1-f$ and $f_2'=f_2-f$, Eq.~(\ref{eq:GNLI_GN_rect}) can be written as:
\begin{equation}
    \begin{aligned}
        &G_{\mathrm{NLI,rect}}(f)  =   A_{\mathrm{rect}}(f)   \\ 
              & \int_{a'}^{b'}     \int_{c'}^{d'}  \left| \int_0^{L_{\mathrm{tot}}} \gamma_{\mathrm{rect}}(z)    p_{\mathrm{rect}} (z) 
             e^{j 4 \pi^2 f_1' f_2' B(z)} dz\right|^2  df_1' df_2'
    \end{aligned}
    \label{eq:GNLI_rect}
\end{equation}
where $a' = a-f, b' = b-f, c'=c-f, d'=d-f,  A_{\mathrm{rect}}(f) =  \frac{16}{27}  \Gamma(f)  G_{\mathrm{WDM,1}}G_{\mathrm{WDM,2}}  G_{\mathrm{WDM},3} $.  The factor $\Gamma(f)$ accounts for the frequency-dependent lumped gain/loss between the transmitter and receiver. Within the rectangle under consideration, the three PSD factors associated with the interacting spectral components are assumed to be constant. They are taken outside the integral, which is key to solving the GN-model integral. This locally constant representation is accurate when the rectangles are sufficiently small, and its error can be controlled through refinement of the partition, particularly near spectral roll-off regions.  $B(z)=\int_{0}^{z} \beta_{2,\mathrm{eff,rect}}(\zeta)\,d\zeta $ is the accumulated  dispersion from the transmitter towards $z$. The quantities $\gamma_{\mathrm{rect}}(z)$, $p_{\mathrm{rect}}(z)$, and $\beta_{2,\mathrm{eff,rect}}(z)$ denote, respectively, the  \textit{effective} nonlinear coefficient, SPP, and dispersion in the rectangle. 
Their definitions follow  Appendix A in \cite{2026_JLT_Poggiolini}, where more details are provided. Here we list their expressions for your reference:
\begin{equation}
    \begin{aligned} 
    & \gamma_{\mathrm{rect}}(z) \approx \frac{2\pi f}{c} \cdot \frac{n_2(z)}{A_{\text{eff}}(z, f, f_1,f_2,f_3)} 
    \end{aligned}
    \label{eq:gamma_approximation}
\end{equation}
where $c$ denotes the speed of light in vacuum and $n_2(z)$ is the nonlinear refractive index of the fiber. The quantity $A_{\text{eff}}(z, f, f_1,f_2,f_3)$ is the cross-effective area of the four frequency components $(f, f_1,f_2,f_3)$ involved in the non-linear Kerr interaction.
\begin{equation}
\begin{aligned}
     \beta&_{2,\mathrm{eff,rect}}(z) = \beta_2(z)
        + \pi \beta_3(z) \bigl(f_1 + f_2 - 2 f_c
        \bigr) \\
      & + \frac{2}{3} \pi^2 \beta_4(z) [
          (f_1 - f_c)^2 + (f_1 - f_c)(f_2 - f_c) \\
      & + (f_2 - f_c)^2 +\frac{1}{2}(f_1 - f)( f_2 - f)]
\end{aligned}
\label{eq:beta2_effective}
\end{equation}
Here, $f_c$ is the reference frequency for the dispersion expansion, and $\beta_2(z)$, $\beta_3(z)$, and $\beta_4(z)$ are the second-, third-, and fourth-order dispersion coefficients, respectively. Fifth- and higher-order terms are neglected.
\begin{equation}
    \begin{aligned}           
    p_{\mathrm{rect}}(z) = \sqrt{\frac{
    p(f_1,z) p(f_2, z) p(f_3, z) }{
    p(f, z)
    }}
    \label{eq:rho_general}
    \end{aligned}
\end{equation}
where $p(f,z)$ is the `normalized SPP' of the channel to which $f$ belongs, relative to the launch power.
These quantities may differ among different rectangles and spans.

\subsection{Span-pair decomposition}

For a link comprising $N_s$ spans, as shown in Fig.~\ref{fig:coherent_NLI}, the NLI fields generated at different positions accumulate coherently at the receiver. Their contributions can be decomposed into two types: SS and XS contributions. The contribution generated within pan $i$, denoted by $(i,i)$ and represented by a blue bracket in Fig.~\ref{fig:coherent_NLI}, is an SS contribution. The coherent interaction generated in two distinct spans $i$ and $j$, denoted by $(i,j), i \neq j$ and represented by a red dashed arrow, is an XS contribution. The SS terms are generally dominant, whereas the XS terms generally act as a correction.
Existing CFMs typically represent the SS contributions accurately, but treat the XS contributions more approximately. Such treatments can be sufficiently accurate when accumulated dispersion rapidly de-correlates the NLI fields generated in different spans. In low-dispersion and/or low-baud-rate systems, however, residual XS coherence may remain significant. In this paper, we therefore retain both contributions and derive them within a unified framework. 

For this link, we define the span boundaries as $ Z_0 = 0,  Z_m = \sum_{\ell=1}^{m} L_{s}^{(\ell)}, m=1,\ldots,N_{s}$,
where $L_{s}^{(\ell)}$ is the length of span $\ell$. Within span $m$, the local coordinate is $z_m = z-Z_{m-1}, 0 \le z_m \le L_s^{(m)}$. Therefore, the NLI generated along the whole link can be decomposed into a coherent sum of span-wise contributions:
\begin{equation}
    \begin{aligned}
        G&_{\mathrm{NLI,rect}}(f)  =  A_{\mathrm{rect}}(f)  \int_{a'}^{b'}  \int_{c'}^{d'} \left|\sum_{m=1}^{N_{s}} \mathcal{I}_{\mathrm{rect}}^{(m)}(f_1',f_2')\right|^2 df_1' df_2' \\
    \end{aligned}
    \label{eq:GNLI_rect_spans_I}
\end{equation}
where span $m$ has:
\begin{equation}
    \begin{aligned}
        \mathcal{I}&_{\mathrm{rect}}^{(m)}  = e^{j4\pi^2 f_1'f_2' B_m} \\
        &\int_{0}^{L_{s}^{(m)}} \mathbf{p}_{\mathrm{rect}}^{(m)}(z) e^{j4\pi^2 f_1'f_2'\beta_{2,\mathrm{eff,rect}}^{(m)} z_m} dz_m
    \end{aligned}
    \label{eq:integral_z}
\end{equation}
where $B_m = \sum_{\ell=1}^{m-1} \beta_{2,\mathrm{eff,rect}}^{(\ell)} L_{s}^{(\ell)} $ is the accumulated dispersion before span $m$. $\mathbf{p}_{\mathrm{rect}}^{(m)}(z)$ is defined as the NLI stimulus profile, or NSP:
\begin{equation}
    \begin{aligned}
        \mathbf{p}_{\mathrm{rect}}^{(m)}(z) = \gamma_{\mathrm{rect}}^{(m)} p_{\mathrm{rect}}^{(m)} (z) 
    \end{aligned}
    \label{eq:NSP}
\end{equation}
The NSP represents the effective strength of nonlinear field generation at position $z$ within span $m$. Physically, the nonlinear contribution generated at each position is determined not only by the local optical power, but also by the nonlinear coefficient. Hence, regions with larger signal power or larger nonlinear coefficient contribute more strongly to the generated NLI field. The NSP $\mathbf{p}_{\mathrm{rect}}^{(m)}(z)$ also has the support $[0, L_{s}^{(m)}]$. 

By expanding the squared modulus, the NLI can be expressed as a sum over all span pairs:
\begin{equation}
    \begin{aligned}
       G&_{\mathrm{NLI,rect}}(f)  =  \sum_{m=1}^{N_s} \sum_{n=1}^{N_s} G_{\mathrm{NLI,rect}}^{(m,n)}(f)
    \end{aligned}
    \label{eq:GNLI_rect_spans_II}
\end{equation}
where:
\begin{equation}
    \begin{aligned}
        G&_{\mathrm{NLI,rect}}^{(m,n)}(f) = A_{\mathrm{rect}}(f) K_{\mathrm{NLI,rect}}^{(m,n)}(f)
\end{aligned}
\label{eq:GNLI_rect_mn}
\end{equation}
with:
\begin{equation}
    \begin{aligned}
        K&_{\mathrm{NLI,rect}}^{(m,n)}(f) = \int_{a'}^{b'} \int_{c'}^{d'} \mu_{\mathrm{rect}}^{(m,n)}(f_1',f_2') df_1' df_2' \\
        \mu&_{\mathrm{rect}}^{(m,n)}(f_1',f_2') =    \mathcal{I}_{\mathrm{rect}}^{(m)}(f_1',f_2') \big[\mathcal{I}_{\mathrm{rect}}^{(n)}(f_1',f_2')\big]^*
    \label{eq:mu_rect_mn}
     \end{aligned}
\end{equation}

Exchanging $m$ and $n$, $K_{\mathrm{rect}}^{(n,m)}(f_1',f_2') = \left[K_{\mathrm{rect}}^{(m,n)}(f_1',f_2')\right]^*$, and hence $G_{\mathrm{NLI,rect}}^{(n,m)}(f') = \left[G_{\mathrm{NLI,rect}}^{(m,n)}(f)\right]^*$. The diagonal terms, $m=n$, correspond to the SS contributions, whereas the off-diagonal terms, $m\neq n$, correspond to the XS contributions. Therefore, Eq.~(\ref{eq:GNLI_rect_spans_II}) can be rewritten as:
\begin{equation}
    G_{\mathrm{NLI,rect}}(f) = G_{\mathrm{NLI,rect}}^{\mathrm{SS}}(f) + G_{\mathrm{NLI,rect}}^{\mathrm{XS}}(f)
    \label{eq:GNLI_rect_SS_XS}
\end{equation}
where:
\begin{equation}
    G_{\mathrm{NLI,rect}}^{\mathrm{SS}}(f) = \sum_{m=1}^{N_s} G_{\mathrm{NLI,rect}}^{(m,m)}(f)
    \label{eq:GNLI_rect_SS}
\end{equation}
and
\begin{equation}
    \begin{aligned}
        G_{\mathrm{NLI,rect}}^{\mathrm{XS}}(f) &= \sum_{\substack{m,n=1\\m\neq n}}^{N_s} G_{\mathrm{NLI,rect}}^{(m,n)}(f) \\
        &=2\sum_{m=1}^{N_s-1} \sum_{n=m+1}^{N_s}  \operatorname{Re} \left\{ G_{\mathrm{NLI,rect}}^{(m,n)}(f)\right\}
        \label{eq:GNLI_rect_XS}
    \end{aligned}    
\end{equation}

The key step in deriving a closed-form or semi-analytical model is therefore the efficient evaluation of the core integral $K_{\mathrm{NLI,rect}}^{(m,n)}(f)$. 
We first derive the formulation for the SS contributions, and then extend the same construction to the XS contributions with $m<n$.

\section{semi-analytical evaluation of single-span contributions}
\label{sect:semi_analytical_SS}
We first consider the SS contributions in Eq.~(\ref{eq:GNLI_rect_SS}). The corresponding core factor, $K_{\mathrm{NLI,rect}}^{(m,m)}(f)$, is obtained by setting $n=m$ in Eq.~(\ref{eq:mu_rect_mn}),
%\begin{equation}
%    \begin{aligned}
%        K&_{\mathrm{NLI,rect}}^{(m,m)}(f)  =  \int_{a'}^{b'}  \int_{c'}^{d'} \mu_{\mathrm{NLI,rect}}^{(m,m)}(f_1', f_2')  df_1' df_2' 
%    \end{aligned}
%    \label{eq:GN_rec_core}
%\end{equation}
where $\mu_{\mathrm{NLI,rect}}^{(m,m)}(f_1', f_2')$ is obtained by substituting Eq.~(\ref{eq:integral_z}) into Eq.~(\ref{eq:mu_rect_mn}):
\begin{equation}
    \begin{aligned}
        \mu&_{\mathrm{NLI,rect}}^{(m,m)}(f_1', f_2')  \\
%            & =  \left| \int_{0}^{L_{s}^{(m)}} \mathbf{p}_{\mathrm{rect}}^{(m)}(z_m) e^{j4\pi^2 f_1'f_2'\beta_{2,\mathrm{eff,rect}}^{(m)} z_m} dz_m \right|^2   \\  
            & = \int_{0}^{L_{s}^{(m)}} \int_{0}^{L_{s}^{(m)}} \mathbf{p}_{\mathrm{rect}}^{(m)}(z_1) \mathbf{p}_{\mathrm{rect}}^{(m)}(z_2) 
            e^{j\Omega(z_1-z_2)} dz_1 dz_2 \\
    \end{aligned}
    \label{eq:GNLI_rect_LK}
\end{equation}
with $\Omega = 4\pi^2 f_1'f_2'\beta_{2,\mathrm{eff,rect}}^{(m)}$.
The core factor $K_{\mathrm{NLI,rect}}^{(m,m)}(f)$ now contains two frequency integrations and two space integrations. We derive two equivalent semi-analytical representations by changing the order in which these integrations are evaluated. In Method 1, the space integrations are evaluated first. In Method 2, the rectangular frequency integration is evaluated first.

\subsection{\textbf{Method 1: z-integral first}}
\label{sect:SS_method1}
We start from Eq.~(\ref{eq:GNLI_rect_LK}) by introducing the spatial change of variables: 
\begin{equation}
    \Sigma = \frac{z_1+z_2}{2}, \quad \Delta=z_1-z_2 
    \label{eq:zz_add_sub}
\end{equation}
The absolute value of the Jacobian is one. The square domain
$(z_1,z_2)\in[0,L_{s}^{(m)}]\times[0,L_{s}^{(m)}]$ is transformed into a diamond-shaped domain in the $(\Sigma,\Delta)$ plane. Thus $-L_{s}^{(m)}\leq \Delta \leq L_{s}^{(m)}$, and $|\Delta|/2 \leq \Sigma \leq L_{s}^{(m)}-|\Delta|/2$ for a given $\Delta$. Consequently, Eq.~(\ref{eq:GNLI_rect_LK}) changes to:
\begin{equation}
    \begin{aligned}
        \mu&_{\mathrm{NLI,rect}}^{(m,m)}(f_1', f_2')  =  \int_{-L_{s}^{(m)}}^{L_{s}^{(m)}} e^{j\Omega\Delta} d\Delta \\
        & \big(\int_{|\Delta|/2}^{L_{s}^{(m)}-|\Delta|/2} \mathbf{p}_{\mathrm{rect}}^{(m)}\left(\Sigma+\frac{\Delta}{2}\right) \mathbf{p}_{\mathrm{rect}}^{(m)}\left(\Sigma-\frac{\Delta}{2}\right)  d\Sigma \big)
    \end{aligned}
    \label{eq:mu_diamond}
\end{equation}

We then extend the NSP outside the span $m$ by simply defining that:
\begin{equation}
    \mathbf{p}_{\mathrm{rect}}^{(m)}(z) = 0, z \notin [0,L_{s}^{(m)}]
\end{equation}
In this way, the inner integral in $\Sigma$ in Eq.~(\ref{eq:mu_diamond})  does not need any limits, as it is automatically generated by the finite support of $\mathbf{p}_{\mathrm{rect}}^{(m)}(z)$ (i.e. $p_{\mathrm{rect}}^{(m)} (z) $). Therefore, this integral can be equivalently written over the whole real axis and identified as the autocorrelation:
%In particular, defining the autocorrelation as:
%\begin{equation}
%    R_{p}(\Delta)  = \int_{-\infty}^{\infty} p \left( y+\frac{\Delta}{2}\right)
%        p\left(y-\frac{\Delta}{2}\right) dy
%    \label{autocorrelation}
%\end{equation}
\begin{equation}
    \begin{aligned}
        &\int_{|\Delta|/2}^{L_{s}^{(m)}-|\Delta|/2} \mathbf{p}_{\mathrm{rect}}^{(m)}\left(\Sigma+\frac{\Delta}{2}\right)        \mathbf{p}_{\mathrm{rect}}^{(m)}\left(\Sigma-\frac{\Delta}{2}\right)  d\Sigma   \\ 
        = & \int_{-\infty}^{\infty} \mathbf{p}_{\mathrm{rect}}^{(m)}\left(\Sigma+\frac{\Delta}{2}\right) \mathbf{p}_{\mathrm{rect}}^{(m)}\left(\Sigma-\frac{\Delta}{2}\right)  d\Sigma  \\
        = &R_{\mathbf{p}}^{(m)}(\Delta)
    \end{aligned}
\label{eq:Rp_autocorrelation}
\end{equation}
The autocorrelation quantifies the spatial overlap of the NSP with a shifted replica of itself. It reaches its maximum at zero lag, $\Delta=0$, where the NSP fully overlaps with itself, and generally decreases as $|\Delta|$ increases because the overlap between the two shifted profiles is progressively reduced.

\begin{figure}
    \centering
    \includegraphics[width=0.9\columnwidth]{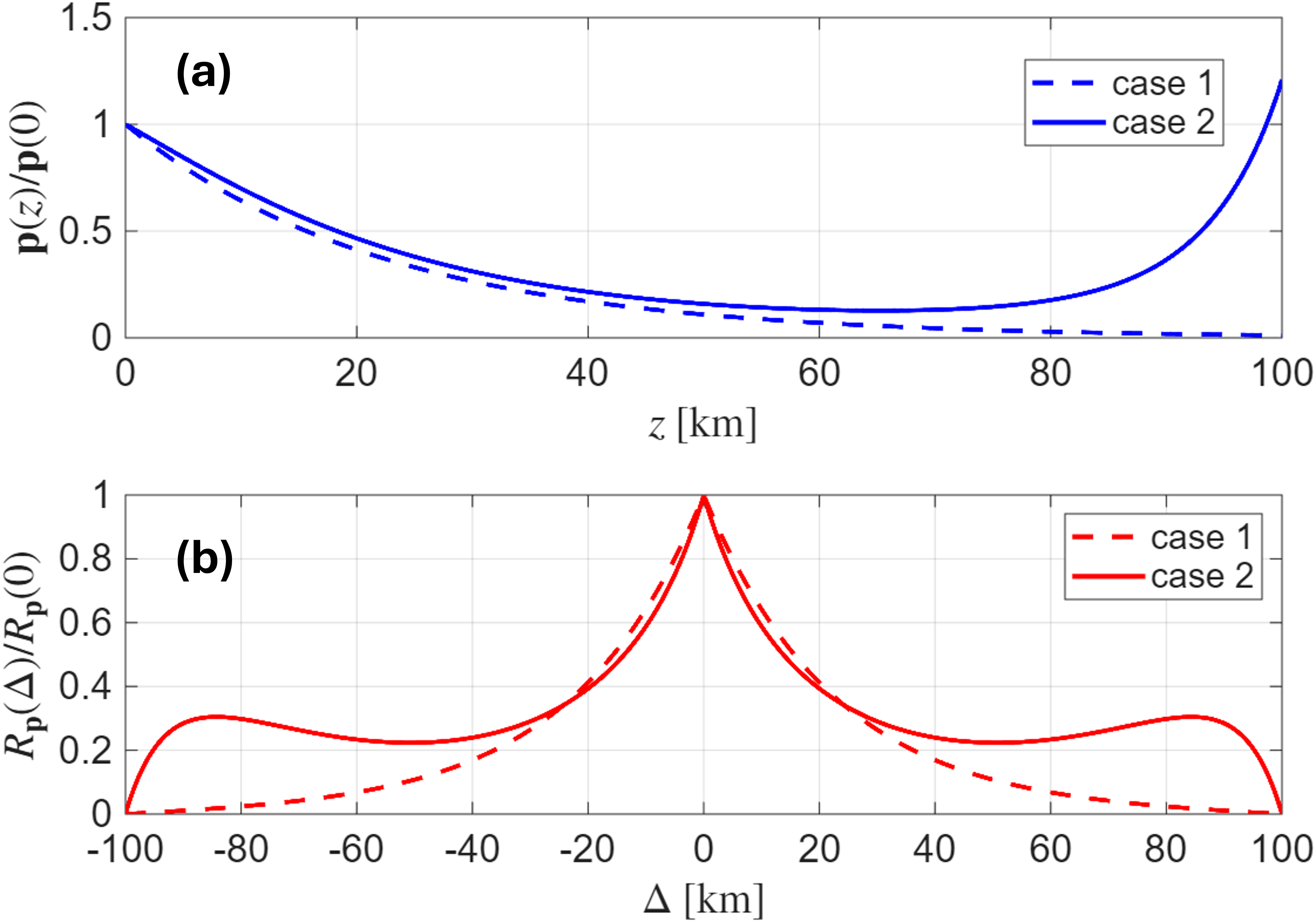}
    \caption{Normalized (a) NSP and (b) its normalized autocorrelation, for case 1 (dashed curves) with an exponential SPP and case 2 (solid curves) with a complicated SPP taken from the 75-th channel in a 150-channel C+L+S system with significant ISRS and backward amplification in Fig.9 of \cite{2026_JLT_Poggiolini}.}
    \label{fig:Rp_2case}
\end{figure} 

\begin{figure}
    \centering
    \includegraphics[width=0.9\columnwidth]{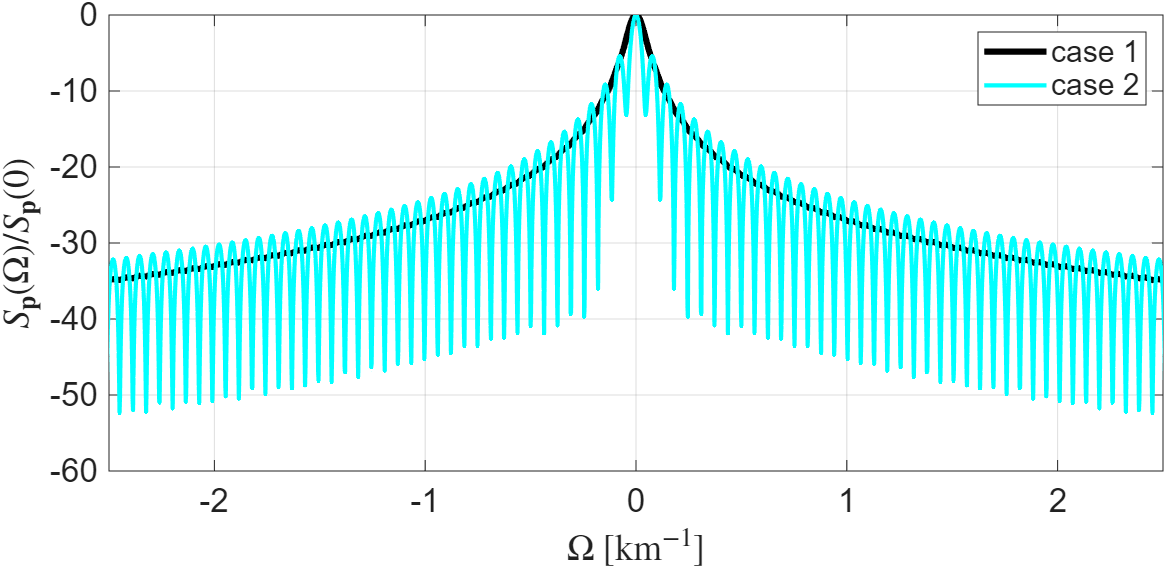}
    \caption{Normalized energy spectrum for case 1 (black) with an exponential SPP and case 2 (cyan) with a complicated SPP taken from the 75-th channel in a 150-channel C+L+S system with significant ISRS and backward amplification in Fig.9 of \cite{2026_JLT_Poggiolini}.}
    \label{fig:Sp_2case}
\end{figure} 

Figure~\ref{fig:Rp_2case} provides two representative NSPs and their autocorrelations. In case 1, shown by dashed curves, the SPP is affected only by the intrinsic fiber attenuation. That means, it follows an exponential decay along the span, i.e., $p(z)=\exp(-2\alpha z)$. Its autocorrelation decreases monotonically with $|\Delta|$. Case 2 (solid curves) contains a more complicated SPP, which is taken from the 75-th channel in a 150-channel C+L+S system (in Fig.~9, \cite{2026_JLT_Poggiolini}) with significant ISRS and backward Raman amplification. Therefore, the SPP bends upward near the fiber output. As expected, the autocorrelation reaches its maximum when the lag $\Delta=0$, and decreases non-monotonically because: When the shift becomes sufficiently large, the high-power region near the span input can partially overlap with the Raman-enhanced high-power region near the span output, producing a secondary increase in the autocorrelation.

Eq.~(\ref{eq:mu_diamond}) is now reduced to:
\begin{equation}
    \begin{aligned}
        \mu&_{\mathrm{NLI,rect}}^{(m,m)}(f_1', f_2') = \int_{-L_{s}^{(m)}}^{L_{s}^{(m)}} R_{\mathbf{p}}^{(m)}(\Delta) e^{j\Omega\Delta} d\Delta 
    \end{aligned}
    \label{eq:mu_Rp}
\end{equation}
Since the support of the NSP is $[0,L_{s}^{(m)}]$, the support of $R_{\mathbf{p}}^{(m)}$ is then $[-L_{s}^{(m)},L_{s}^{(m)}]$. Consequently, Eq.~(\ref{eq:mu_Rp}) can be equivalently written as:
\begin{equation}
    \begin{aligned}
         \mu_{\mathrm{NLI,rect}}^{(m,m)}(f_1', f_2') & = \int_{-\infty}^{\infty}  R_{\mathbf{p}}^{(m)}(\Delta) e^{j\Omega\Delta} d\Delta  = S_{\mathbf{p}}^{(m)}(\Omega)
    \end{aligned}
    \label{eq:mu_Sp}
\end{equation}
because the energy spectrum $S_{\mathbf{p}}^{(m)}(\Omega)$ is defined as the Fourier transform of its autocorrelation. It describes its response to the phase mismatch $\Omega$. The quantity $\Omega$ is $z$-independent, but depends on chromatic dispersion and frequencies. $S_{\mathbf{p}}^{(m)}(\Omega)$ is an even function, satisfying:
\begin{equation}
    S_{\mathbf{p}}^{(m)}(\Omega) = S_{\mathbf{p}}^{(m)}(-\Omega)
    \label{eq:Sp}
\end{equation}
The energy spectrum reaches its maximum when the phase mismatching term $\Omega=0$, corresponding to perfect phase matching and fully coherent accumulation along the span. As $\Omega$ increases, the phase mismatch induces faster oscillations along $z$, leading to destructive cancellation of nonlinear contributions and, consequently, a reduced NLI generation efficiency. 

Figure~\ref{fig:Sp_2case} shows the normalized energy spectrum of the two cases in Fig.~\ref{fig:Rp_2case}. The value is at least 30~dB below its maximum value at $\Omega= 0$ when $|\Omega| \gtrsim 2~\mathrm{km}^{-1}$ in both cases. Hence, a rectangle whose phase-mismatch term lies mainly in this high $|\Omega|$ region can be safely neglected, since its contribution to the NLI is strongly suppressed by phase mismatch. In case 1, the energy spectrum shows a smooth decay away from $\Omega= 0$, approximately following a Lorentzian-like dependence since $\exp(-2\alpha L_s)$ is small. However, case 2 has many peaks and nulls, reflecting constructive and destructive interference of nonlinear fields generated in separated high-power regions along the span. The approximate oscillation period is governed by the position of the side peak $\delta_R$ in $R_{\mathbf{p}}^{(m)}(\Delta)$, with $\Delta \Omega \sim 2 \pi/ \delta_R$.  In case 2, \(\delta_R \approx 85~\mathrm{km}\), giving 
\(\Delta \Omega \sim 0.074~\mathrm{km}^{-1}\).

We then obtain: 
\begin{equation}
    \begin{aligned}
        K_{\mathrm{NLI,rect}}^{(m,m)}(f)  =  \int_{a'}^{b'}  \int_{c'}^{d'} S_{\mathbf{p}}^{(m)}\big( B_2^{(m)} f_1' f_2'  \big) df_1' df_2'
    \end{aligned}
    \label{eq:f1f2_integral}
\end{equation}
where $B_2^{(m)} = 4\pi^2\beta_{2,\mathrm{eff,rect}}^{(m)}$. The rectangle may be fully contained within a single quadrant, as in the case of MCI islands. It may cross one coordinate axis, as for XCI islands, or it may intersect both coordinate axes, as for the SCI island. These different geometric configurations make a direct derivation cumbersome. 

Nevertheless, because the energy spectrum is even with respect to $f_1'\cdot f_2'$, all quadrant configurations can be mapped to a first-quadrant primitive. The contribution over an arbitrary rectangle can then be reconstructed by symmetry. Therefore, we only need to consider a rectangular domain located entirely in the first quadrant, satisfying $0<a'<b', 0<c'<d'$. 

 We then introduce the change of variables:
\begin{equation}
    \xi = f_1'f_2',     \quad  \eta = f_1' 
    \label{eq:app_xi_eta}
\end{equation}
Their intervals are:
\begin{equation}
    \begin{aligned}
        & \xi_{\text{min}}\le \xi \le \xi_{\text{max}}   \\ 
        & \eta_{\text{min}} \le \eta  \le  \eta_{\text{max}}
    \label{eq:interval}
    \end{aligned}   
\end{equation}
where:
\begin{equation}
    \begin{aligned}
        & \xi_{\text{min}} = a'c', \xi_{\text{max}}=b'd'\\
        & \eta_{\text{min}} = \max\left(a',{\xi}/{d'}\right) \\
        & \eta_{\text{max}} = \min\left(b',{\xi}/{c'}\right)
    \label{eq:interval_limits}
    \end{aligned}   
\end{equation}
Therefore, Eq.~(\ref{eq:f1f2_integral}) becomes:
\begin{equation}
    K_{\mathrm{NLI,rect}}^{(m,m)}(f)   =  \int_{\xi_{\text{min}} }^{\xi_{\text{max}}} S_{\mathbf{p}}^{(m)}\big( B_2^{(m)} \xi  \big)
    \big(\int_{\eta_{\text{min}}}^{\eta_{\text{max}}}\frac{d\eta}{\eta}\big) d\xi 
    \label{eq:KNLI_rect_single_integral_pre_weight}
\end{equation}

Defining the weight:
\begin{equation}
    \begin{aligned}
        W_{\mathrm{rect}}(\xi) = \int_{\eta_{\text{min}}}^{\eta_{\text{max}}} \frac{d\eta}{\eta}
         =\ln\left(\frac{\eta_{\text{max}} }{\eta_{\text{min}}} \right)
    \end{aligned}  
    \label{eq:app_rect_weight_definition}
\end{equation}
we obtain the final expression with a single integral as:
\begin{equation}
    K_{\mathrm{NLI,rect}}^{(m,m)}(f) = \int_{a'c'}^{b'd'} S_{\mathbf{p}}^{(m)}\big( B_2^{(m)} \xi  \big)
    W_{\mathrm{rect}}(\xi) d\xi    
    \label{eq:solution_single_span}
\end{equation}
This gives the final expression of the SS contribution from span $m$. If one of the lower limits lies on an axis, i.e., $a'=0$ or $c'=0$, the same single integral remains valid. But $\xi_{\min}=0$, and the weight function \(W_{\mathrm{rect}}(\xi)\) changes. The logarithmic singularity that may appear at $\xi=0$ is integrable and therefore does not cause any difficulty in the evaluation of the integral.

As a representative primitive case, a rectangle used for SCI evaluation in the first-quadrant rectangle whose lower-left corner coincides with the origin:
\begin{equation}
    a'=0, \quad  b'>0, \quad c'=0, \quad    d'>0 
    \label{eq:app_origin_corner_rectangle}
\end{equation}
Eq.~(\ref{eq:interval_limits}) becomes:
\begin{equation}
    \begin{aligned}
        & \xi_{\text{min}} = 0, \xi_{\text{max}} = b'd'\\
        & \eta_{\text{min}} = {\xi}/{d'}, \eta_{\text{max}} = b'
        \label{eq:SCI_interval_limits}
    \end{aligned}   
\end{equation}
And the weight function can be explicitly written as:
\begin{equation}
        W_{\mathrm{rect}}(\xi) = \ln\left(\frac{b'd'}{\xi}\right)
\end{equation}
Thus:
\begin{equation}
        K_{\mathrm{SCI,rect}}^{(m,m)}(f) = \int_{0}^{b'd'} S_{\mathbf{p}}^{(m)} \left(B_2^{(m)}\xi\right) 
            \ln\left(\frac{b'd'}{\xi}\right) d\xi
\end{equation}
The logarithmic singularity at $\xi=0$ is integrable. Therefore, the expression is well defined provided that
$S_{\mathbf{p}}^{(m)}(B_2^{(m)}\xi)$ remains finite in the neighborhood of $\xi=0$.

\subsection{\textbf{Method 2: f-integral first}}
\label{sect:SS_method2}
After substituting Eq.~(\ref{eq:GNLI_rect_LK}) into the core factor $K_{\mathrm{NLI,rect}}^{(m,m)}(f)$, and changing the order of the integration, we obtain:
\begin{equation}
    \begin{aligned}
          K_{\mathrm{NLI,rect}}^{(m,m)}(f)  & =  \int_{0}^{L_{s}^{(m)}} \int_{0}^{L_{s}^{(m)}} \mathbf{p}_{\mathrm{rect}}^{(m)}(z_1) \mathbf{p}_{\mathrm{rect}}^{(m)} (z_2) \\
          &\mathcal{F_{\text{rect}}}(\mathcal{B}^{(m)}) dz_1 dz_2
    \end{aligned}
    \label{eq:integral_method2}
\end{equation}
where $\mathcal{B}^{(m)} = B_2^{(m)}(z_1-z_2) $, and:
\begin{equation}
    \begin{aligned}
        \mathcal{F}_{\text{rect}}(\mathcal{B}) = \int_{a'}^{b'} \int_{c'}^{d'} e^{j\mathcal{B} f_1' f_2'} df_1' df_2' 
    \end{aligned}
    \label{eq:FB_definition}
\end{equation}
which represents the coherence response function of a rectangular integration island with respect to the accumulated dispersion difference $\mathcal{B}$.  It satisfies the Hermitian symmetry:
\begin{equation}
    \mathcal{F}_{\mathrm{rect}}(-\mathcal{B}) =   \mathcal{F}_{\mathrm{rect}}^{*}(\mathcal{B})
\end{equation}
Since changing $z_1-z_2$ to $z_2-z_1$ changes $\mathcal{B}$ to $-\mathcal{B}$, the contributions from the two symmetric regions $z_1>z_2$ and $z_1<z_2$ are complex conjugates. Consequently, the imaginary parts cancel in the self-span contribution $K_{\mathrm{NLI,rect}}^{(m,m)}(f)$, and only the real part of $\mathcal{F}_{\mathrm{rect}}(\mathcal{B})$ contributes after integration over $z_1$ and $z_2$.

When $\mathcal{B} = 0$, $\mathcal{F}_{\text{rect}}(\mathcal{B}) = (b'-a')(d'-c')$, giving the maximum coherent contribution. This condition is satisfied either for $z_1=z_2$, where the two NSP profiles are superimposed without spatial lag, or for $\beta_{2,\mathrm{eff,rect}}^{(m)}=0$, where there is no dispersion-induced phase mismatch. 
As $|\mathcal{B}|$ increases, the phase varies more rapidly over the rectangle and the coherent contribution decreases. The de-correlation rate depends strongly on the location of the rectangle in the $(f_1',f_2')$ plane. SCI-like rectangles containing the origin retain 
low-\(|f_1'f_2'|\) regions and decorrelate slowly, XCI-like rectangles crossing one axis show intermediate behavior, and MCI-like rectangles away from both axes decorrelate most rapidly. The oscillation scale is governed by the spread of the product variable$f_1'f_2'$.

When $\mathcal{B}\neq 0$, we first integrate over \(f_2'\) as:
\begin{align}
    \int_{c'}^{d'} e^{j\mathcal{B}f_1'f_2'} df_2'= \frac{ e^{j\mathcal{B}f_1'd'} - e^{j\mathcal{B}f_1'c'}}{j\mathcal{B}f_1'} 
    \label{eq:f2_integral}
\end{align}
Strictly speaking, it contains a removable singularity at $f_1'=0$. To express the remaining integral, define the entire function
\begin{equation}
    \mathcal{E}(z)
    =
    \sum_{k=1}^{\infty}\frac{z^k}{k\,k!},
    \qquad \mathcal{E}(0)=0.
\end{equation}
Since $d\mathcal{E}(z)/dz=(e^z-1)/z$, Substituting \eqref{eq:f2_integral} into the remaining integral over $f_1'$ gives:
\begin{equation}
    \begin{aligned}
    \mathcal{F}_{\text{rect}}(\mathcal{B}) &= \frac{1}{j\mathcal{B}} \int_{a'}^{b'} \frac{ e^{j\mathcal{B}d'f_1'} - e^{j\mathcal{B}c'f_1'}} {f_1'} df_1' \\
     &= \frac{1}{j\mathcal{B}} \Big[\mathcal{E} \left(j\mathcal{B}b'd'\right)
            -\mathcal{E} \left(j\mathcal{B}a'd'\right) \\
            &-\mathcal{E}\!\left(j\mathcal{B}b'c'\right) +\mathcal{E}\!\left(j\mathcal{B}a'c'\right)\Big] 
    \end{aligned}  
    \label{eq:FB_Ei}
\end{equation}
When one or more arguments vanish, or when $|\mathcal{B}|$ is small, the original
integral in Eq.~(\ref{eq:FB_definition}) converges to:
\begin{equation}
    \begin{aligned}
    \mathcal{F}_{\mathrm{rect}}(\mathcal{B}) &= \sum_{k=0}^{\infty}
    \frac{(j\mathcal{B})^k}{k!(k+1)^2} \\
         & \times \left((b')^{k+1}-(a')^{k+1}\right) \left((d')^{k+1}-(c')^{k+1}\right)
    \end{aligned}
    \label{eq:FB_series}
\end{equation}
providing a numerically stable evaluation. The $k=0$ term recovers $\mathcal{F}_{\text{rect}}(\mathcal{B})= (b'-a')(d'-c')$.

\begin{figure}
    \centering
    \includegraphics[width=0.9\columnwidth]{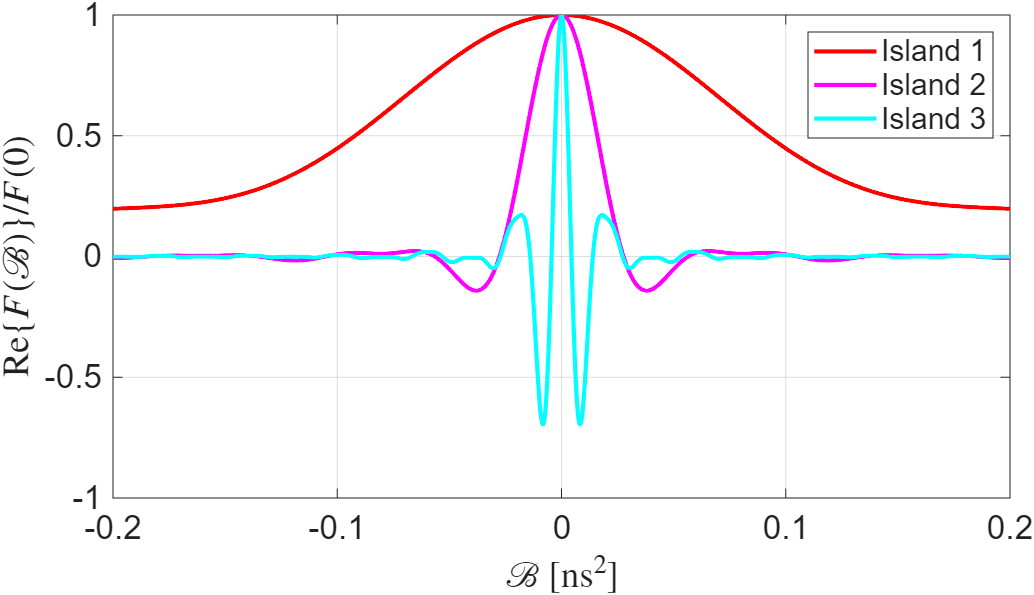}
    \caption{Normalized coherence response function of the three representative integration islands. Island 1, Island 2, and Island 3 correspond to the SCI-, XCI-, and MCI-like islands, respectively.}
    \label{fig:FB_3island}
\end{figure}

Figure~\ref{fig:FB_3island} provides the normalized coherence response function of the three representative integration islands, each with a size of 2~GHz $\times$ 2~GHz:
\begin{itemize}
    \item \textbf{Island 1}: an SCI-like rectangle, defined by \\
           \[
            \begin{gathered}
            a' = -1~\text{GHz},\qquad b' = +1~\text{GHz},\\
            c' = -1~\text{GHz},\qquad d' = +1~\text{GHz}.
            \end{gathered}
            \]
    \item \textbf{Island 2}: an XCI-like rectangle, defined by \\
            \[
             \begin{gathered}
            a' = -1~\text{GHz},\qquad b' = +1~\text{GHz},\\
            c' = +2~\text{GHz}, \qquad d' = +4~\text{GHz}.
            \end{gathered}
            \]
    \item \textbf{Island 3}: an MCI-like rectangle, defined by \\
            \[
             \begin{aligned}
            a' = +2~\text{GHz},\qquad b' = +4~\text{GHz},\\
            c' = +2~\text{GHz}, \qquad d' = +4~\text{GHz}.
            \end{aligned}
            \]
\end{itemize}

Island 1 exhibits a broad, positive response and the slowest decorrelation, since it contains many frequency pairs with small $|f_1'f_2'|$. In contrast, Island 2 crosses one axis, leading to stronger cancellation, zero crossings, and negative side lobes. Island 3 is displaced from both axes, making $|f_1'f_2'|$ relatively large over the whole integration domain. Therefore, even small variations in $\mathcal{B}$ induce rapid phase rotation, and the fastest oscillation. The negative values correspond to destructive coherent addition, not negative NLI power.

As $\mathcal{F}_{\text{rect}}(\mathcal{B})$ admits a closed-form expression, and $\mathcal{B}$ depends only on the spatial separation $z_1-z_2$, Eq.~(\ref{eq:integral_method2}) can be finally reduced to a semi-analytical expression:

\begin{equation}
    \begin{aligned}
          K_{\mathrm{NLI,rect}}^{(m,m)}(f)  & =  \int_{-L_{s}^{(m)}}^{L_{s}^{(m)}} R_{\mathbf{p}}^{(m)}(\Delta) \mathcal{F}_{\text{rect}}( B_2^{(m)}\Delta) d\Delta 
    \end{aligned}
    \label{eq:SA_method2}
\end{equation}

Equations~(\ref{eq:solution_single_span}) and (\ref{eq:SA_method2}) are equivalent semi-analytical representations of the SS contribution. The preferable representation depends on whether the NSP energy spectrum or its spatial autocorrelation is
more convenient to compute and reuse. The next section extends both representations to XS contributions.

\section{semi-analytical evaluation of cross-span contributions}
\label{sect:semi_analytical_XS}
We now consider the XS contributions, which describe the coherent interference between NLI fields generated in distinct spans. The core factor $K_{\mathrm{NLI,rect}}^{(m,n)}(f)$ is shown in Eq.~(\ref{eq:mu_rect_mn}), with:
\begin{equation}
    \begin{aligned}
        \mu&_{\mathrm{rect}}^{(m,n)}(f_1',f_2') =  e^{j 4\pi^2 f_1'f_2'(B_m-B_n)}\int_{0}^{L_s^{(m)}}\int_{0}^{L_s^{(n)}}\\ 
            &
            \mathbf{p}_{\mathrm{rect}}^{(m)}(z_m) \mathbf{p}_{\mathrm{rect}}^{(n)}(z_n)
             e^{j f_1'f_2' \big(B_2^{(m)}z_m -B_2^{(n)}z_n \big)}dz_n dz_m
    \end{aligned}
    \label{eq:mu_XS_double_integral}
\end{equation}
As discussed in Sect.~\ref{sect:NLI_accumulation}, it is sufficient to evaluate for only \(m<n\). Unlike the SS contributions, the two NSPs may differ because the span parameters and power evolutions can vary along the link. The phase factor with \((B_m-B_n)\) accounts for the relative accumulated dispersion between the two spans. Again, the integral can be evaluated in two ways: by first evaluating the space integrations, or by first evaluating the frequency integrations.

\subsection{\textbf{Method 1: z-integral first}}
\label{sect:XS_method1}
We evaluate the XS contribution using the correlation-based procedure developed for the SS contribution. In the SS case, the longitudinal phase depends only on the spatial separation between two positions within the same span. For two distinct spans, however, the corresponding phase term is $B_2^{(m)}z_m -B_2^{(n)}z_n $, which cannot generally be expressed in terms of the physical spatial separation.

To recover a difference-coordinate representation, we introduce the signed dispersion coordinates $u_q=B_2^{(q)}z_q, q\in{m,n}$, This transformation gives $B_2^{(m)}z_m -B_2^{(n)}z_n =u_m-u_n$, irrespective of the signs and magnitudes of the two dispersion parameters. We assume that $B_2^{(q)}\neq0$. The zero-dispersion cases are then treated separately later.

Next, we define the dispersion-domain NSP as:
\begin{equation}
    \widetilde{\mathbf p}_{\mathrm{rect}}^{(q)}(u) =
        \begin{cases}
        \dfrac{1}{\left|B_2^{(q)}\right|}
        \mathbf p_{\mathrm{rect}}^{(q)}
        \left(\dfrac{u}{B_2^{(q)}}\right),
        & u\in[u_q^-, u_q^+],\\[2mm]
        0, & \text{otherwise}.
        \end{cases}
    \label{eq:dispersion_domain_NSP}
\end{equation}
The factor $1/\lvert B_2^{(q)}\rvert$ accounts for the Jacobian of the coordinate transformation, $u_q^-=\min\left(0,B_2^{(q)}L_s^{(q)}\right), u_q^+=\max \left(0,B_2^{(q)}L_s^{(q)}\right)$. Hence:
\begin{equation}
    \int_0^{L_s^{(q)}}
    \mathbf p_{\mathrm{rect}}^{(q)}(z_q)
    e^{jB_2^{(q)}\xi z_q}\,dz_q
    =
    \int_{-\infty}^{+\infty}
    \widetilde{\mathbf p}_{\mathrm{rect}}^{(q)}(u_q)
    e^{j\xi u_q}\,du_q 
    \label{eq:dispersion_domain_field}
\end{equation}
and Eq.~\eqref{eq:mu_XS_double_integral} becomes:
\begin{equation}
    \begin{aligned}
    \mu_{\mathrm{rect}}^{(m,n)}(\xi)
    &=
    e^{j4\pi^2\xi(B_m-B_n)}\int_{-\infty}^{+\infty}
    \int_{-\infty}^{+\infty} \\
    &    \widetilde{\mathbf p}_{\mathrm{rect}}^{(m)}(u_m)
    \widetilde{\mathbf p}_{\mathrm{rect}}^{(n)}(u_n)
    e^{j\xi(u_m-u_n)} du_n du_m 
    \end{aligned}
    \label{eq:mu_XS_dispersion_domain}
\end{equation}

Following the same procedure in Sect.~\ref{sect:SS_method1}, we introduce:
\begin{equation}
    \Sigma_u=\frac{u_m+u_n}{2},
    \qquad
    \Delta_u=u_m-u_n.
    \label{eq:deltau}    
\end{equation}
and obtain the cross-correlation of the two dispersion-domain NSPs:
\begin{equation}
    \begin{aligned}
    &\int_{-\infty}^{+\infty} \widetilde{\mathbf p}_{\mathrm{rect}}^{(m)}
    \left(\Sigma_u+\frac{\Delta_u}{2}\right)
    \widetilde{\mathbf p}_{\mathrm{rect}}^{(n)}
    \left(\Sigma_u-\frac{\Delta_u}{2}\right)
    \,d\Sigma_u \\
    =  &R_{\widetilde{\mathbf p}}^{(m,n)}(\Delta_u)
    \end{aligned}
    \label{eq:Rp_cross_dispersion}
\end{equation}
For \(m=n\), Eq.~\eqref{eq:Rp_cross_dispersion} reduces to an even autocorrelation function. For $m\neq n$, the cross-correlation is generally not even because the two dispersion-domain NSPs may have different shapes and supports. Since the NSPs are real-valued, exchanging their order is equivalent to reversing the lag, $ R_{\widetilde{\mathbf p}}^{(m,n)}(-\Delta_u) =  R_{\widetilde{\mathbf p}}^{(n,m)}(\Delta_u).$

\begin{figure}[t]
\centering
\includegraphics[width=0.9\linewidth]{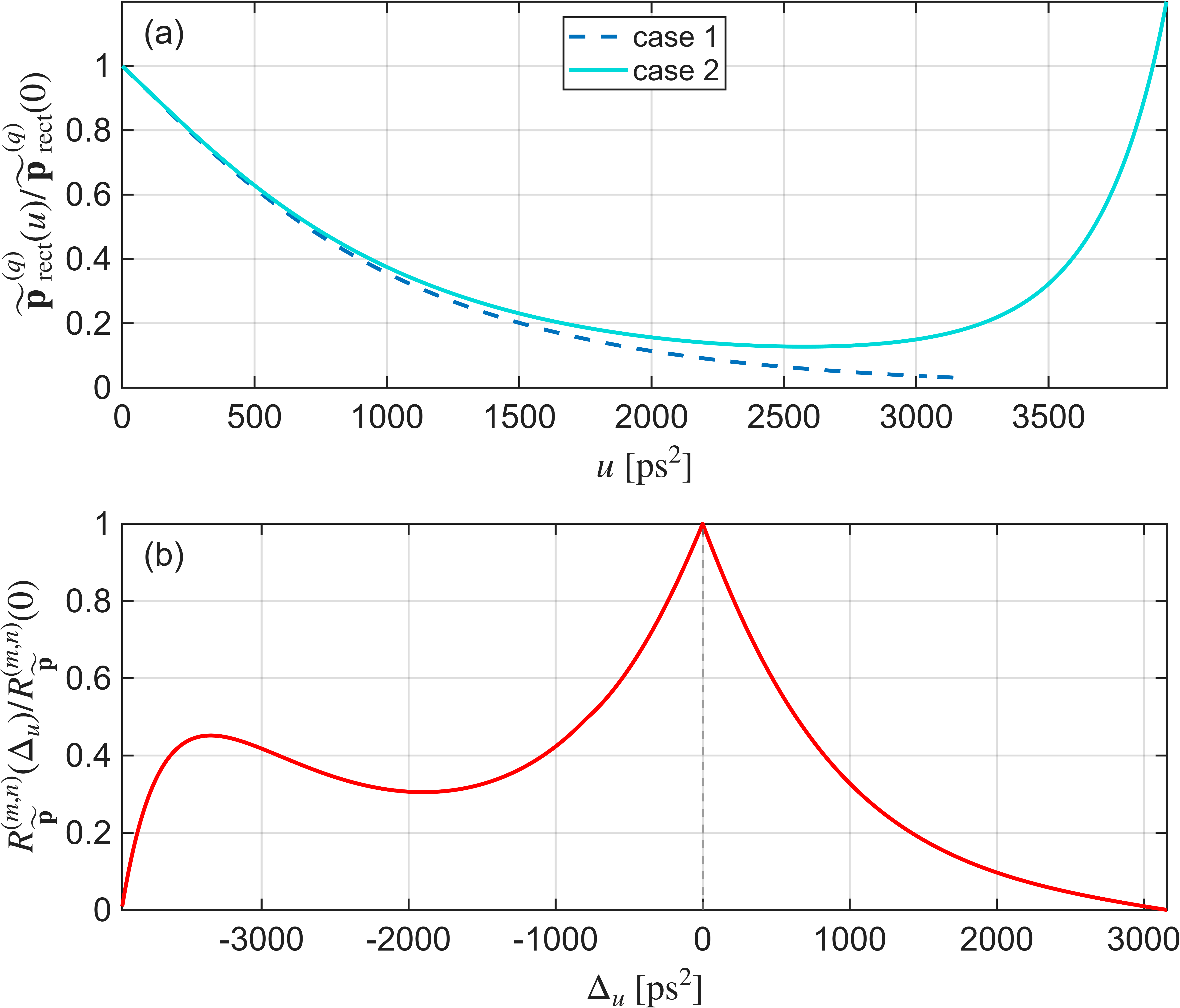}
\caption{Normalized (a) dispersion-domain NSPs, and (b) their normalized ordered cross-correlation \(R_{\widetilde{\mathbf p}}^{(m,n)}(\Delta_u)\), using case 1 (dashed curves) and case 2 (solid curves) introduced in Fig.~\ref{fig:Rp_2case}, with the case 1 span length reduced to 80 km.}
\label{fig:Rp_cross_2case}
\end{figure}

Figure~\ref{fig:Rp_cross_2case} illustrates two distinct dispersion-domain NSPs and their ordered cross-correlation. The two profiles are based on case~1 and case~2 introduced in Fig.~\ref{fig:Rp_2case}, with the span length in case~1 reduced to 80~km for this XS example. The cross-correlation, \(R_{\widetilde{\mathbf p}}^{(m,n)}(\Delta_u)\), quantifies the overlap between the two dispersion-domain profiles as a function of the dispersion-domain lag \(\Delta_u\). It is generally asymmetric about \(\Delta_u=0\), because the two profiles have different shapes and supports.

Equation~\eqref{eq:mu_XS_dispersion_domain} can then be written as:
\begin{equation}
    \mu_{\mathrm{rect}}^{(m,n)}(\xi)
    =
    e^{j4\pi^2\xi(B_m-B_n)}
    \int_{-\infty}^{+\infty}
    R_{\widetilde{\mathbf p}}^{(m,n)}(\Delta_u)
    e^{j\xi\Delta_u}\,d\Delta_u
    \label{eq:mu_XS_Rp}
\end{equation}
Since the cross-spectrum is defined as the Fourier transform of their cross-correlation:
\begin{equation}
    S_{\widetilde{\mathbf p}}^{(m,n)}(\xi)
    =
    \int_{-\infty}^{+\infty}
    R_{\widetilde{\mathbf p}}^{(m,n)}(\Delta_u)
    e^{j\xi\Delta_u}\,d\Delta_u 
    \label{eq:Sp_XS_ordered}
\end{equation}
Defining the complete XS integrand as
\begin{equation}
    \Phi^{(m,n)}(\xi)
    =
    e^{j4\pi^2\xi(B_m-B_n)}
    S_{\widetilde{\mathbf p}}^{(m,n)}(\xi),
\end{equation}
Consequently, we obtain:
\begin{equation}
    K_{\mathrm{NLI,rect}}^{(m,n)}(f) =
    \int_{a'}^{b'}
    \int_{c'}^{d'}
    \Phi^{(m,n)}(\xi)
    df_1'df_2'
    \label{eq:mu_XS_spectrum}
\end{equation}
Unlike the SS energy spectrum, \(S_{\widetilde{\mathbf p}}^{(m,n)}(\xi)\) is generally complex because \(R_{\widetilde{\mathbf p}}^{(m,n)}(\Delta_u)\) is generally not even. Nevertheless, since the cross-correlation is real-valued, its cross-spectrum satisfies $S_{\widetilde{\mathbf p}}^{(m,n)}(-\xi)=\left[S_{\widetilde{\mathbf p}}^{(m,n)}(\xi)\right]^*$. The integrand $\Phi^{(m,n)}(\xi)$, inside $K_{\mathrm{NLI,rect}}^{(m,n)}(f)$, follows the same property. Hence, $\operatorname{Re} \{\Phi^{(m,n)}(\xi)\}$ is even, whereas $\operatorname{Im} \{\Phi^{(m,n)}(\xi)\}$ is odd.

Because the physical contribution of the span pair $\{m,n\}$ is $\operatorname{Re} \{ K_{\mathrm{NLI,rect}}^{(m,n)}(f) \}$, the quadrant-mapping and frequency-product transformation introduced in Sect.~\ref{sect:SS_method1} can be applied to the real part of the XS integrand. Therefore, for a rectangle lying entirely in the first quadrant, we obtain:
\begin{equation}
    \begin{aligned}
    K_{\mathrm{NLI,rect}}^{(m,n)}(f)
    =
    \int_{a'c'}^{b'd'}
    &e^{j4\pi^2\xi(B_m-B_n)}S_{\widetilde{\mathbf p}}^{(m,n)}(\xi)W_{\mathrm{rect}}(\xi)\,d\xi
    \end{aligned}
    \label{eq:K_XS_ordered_single_integral}
\end{equation}

\subsubsection{\textbf{zero-dispersion cases}}
\label{sect:XS_zero_dispersion}

The dispersion-domain transformation requires nonzero local dispersion in both interacting spans. If either $B_2^{(m)}$ or $B_2^{(n)}$ is zero, the corresponding dispersion coordinate collapses to a single point, so the cross-correlation construction in Eqs.~(\ref{eq:dispersion_domain_NSP})--(\ref{eq:mu_XS_Rp}) cannot be used directly. The space integrals in Eq.~(\ref{eq:mu_XS_double_integral}), however, remain separable. Defining:
\begin{equation}
    H_q(\xi)=
    \int_0^{L_s^{(q)}}\mathbf p_{\mathrm{rect}}^{(q)}(z)
    e^{jB_2^{(q)}\xi z}\,dz,\qquad q\in\{m,n\}
\end{equation}
The space integrals in Eq.~(\ref{eq:mu_XS_double_integral}) therefore factorize as \(H_m(\xi)H_n^*(\xi)\). The single-integral expression in Eq.~(\ref{eq:K_XS_ordered_single_integral}) remains applicable after replacing $S_{\widetilde{\mathbf p}}^{(m,n)}(\xi)$ by $H_m(\xi)H_n^*(\xi)$.  Specifically, define the integrated NSP of span $m$ as:
\begin{equation}
    A_q=
    \int_0^{L_s^{(q)}}
    \mathbf p_{\mathrm{rect}}^{(q)}(z_q)\,dz_q,\quad q\in\{m,n\}
    \label{eq:XS_integrated_NSP}
\end{equation}
When \(B_2^{(m)}=0\), this product becomes \(A_mH_n^*(\xi)\); when \(B_2^{(n)}=0\), it becomes \(H_m(\xi)A_n\). If both local dispersions vanish, the contribution becomes:
\begin{equation}
    K_{\mathrm{NLI,rect}}^{(m,n)}(f)    =
    A_mA_n\int_{a'c'}^{b'd'}
    e^{j4\pi^2\xi(B_m-B_n)}
    W_{\mathrm{rect}}(\xi)\,d\xi
    \label{eq:XS_both_local_zero}
\end{equation}
The accumulated-dispersion difference $B_m-B_n$ can still be nonzero if other spans are dispersive. In the special case of an entirely dispersion-free link, $B_m=B_n=0$, and the span-pair contribution further reduces to:
\begin{equation}
    K_{\mathrm{NLI,rect}}^{(m,n)}(f) =
    A_mA_n(b'-a')(d'-c')
    \label{eq:XS_entire_link_zero}
\end{equation}
Each span requires only one independent space integration, after which the coherently accumulated contribution is obtained directly.

\subsubsection{\textbf{NLI PSD in homogeneous links}}
We finally consider a link with $N_s$ identical spans, for which the contribution of each span pair depends only on the span separation. Specifically, we assume $L_s^{(q)}=L_s, \mathbf p_{\mathrm{rect}}^{(q)}(z) =\mathbf p_{\mathrm{rect}}(z), \beta_{2,\mathrm{eff,rect}}^{(q)}=\beta_2, B_2^{(q)}=B_2$.
Consider two spans $m$ and $n=m+r$, where $r$ denotes their separation. In Eq.~(\ref{eq:K_XS_ordered_single_integral}), the accumulated-dispersion difference is:
\begin{equation}
    4\pi^2(B_m-B_{m+r})=-rB_2L_s
    \label{eq:identical_span_phase_difference}
\end{equation}
The cross-spectrum reduces to the SS NSP energy spectrum:
\begin{equation}
    S_{\widetilde{\mathbf p}}^{(m,m+r)}(\xi) =  S_{\mathbf p}(B_2\xi)
    \label{eq:identical_span_ordered_spectrum}
\end{equation}
Taking Eqs.~\eqref{eq:K_XS_ordered_single_integral}, \eqref{eq:identical_span_phase_difference}, and \eqref{eq:identical_span_ordered_spectrum} back to  Eq. \eqref{eq:GNLI_rect_mn} gives:
\begin{equation}
    \begin{aligned}
        G&_{\mathrm{NLI,rect}}^{(r)}(f) \triangleq G_{\mathrm{NLI,rect}}^{(m,n)}(f)  \\
        &= A_{\mathrm{rect}}(f)
        \int_{a'c'}^{b'd'}  e^{-jrB_2L_s\xi}S_{{\mathbf p}}(B_2\xi)W_{\mathrm{rect}}(\xi)\,d\xi
\end{aligned}
\label{eq:GNLI_rect_r}
\end{equation}
The case $r=0$ corresponds to $n=m$ and therefore recovers the SS contribution, whereas $r\geq1$ describes the XS contributions between distinct spans. For a link comprising \(N_s\) identical spans, all span pairs separated by \(r\) provide the same contribution, and the number of such pairs is \(N_s-r\). 

The total NLI PSD, including both SS (Eq.~(\ref{eq:GNLI_rect_SS})) and XS contributions (Eq.~(\ref{eq:GNLI_rect_XS})), is consequently:
\begin{equation}
    \begin{aligned}
        G_{\mathrm{NLI,rect}}(f) &= N_s G_{\mathrm{NLI,rect}}^{(0)}(f)  \\
        & +2\sum_{r=1}^{N_s-1} (N_s-r)\operatorname{Re} 
        \left\{ G_{\mathrm{NLI,rect}}^{(r)}(f)\right\}
        \label{eq:GNLI_XS_identical_span}
    \end{aligned}    
\end{equation}
While $S_{{\mathbf p}}(\xi)$ and $W_{\mathrm{rect}}(\xi)$ in Eq.~(\ref{eq:GNLI_rect_r}) are real-valued, the operator ``$\operatorname{Re}$" acts only on the phase term, therefore: 
\begin{equation}
    \begin{aligned}
   G_{\mathrm{NLI,rect}}(f) = &A_{\mathrm{rect}}(f)  \int_{a'c'}^{b'd'}S_{\mathbf p}(B_2\xi) W_{\mathrm{rect}}(\xi) \\   
     & \left(N_s + 2\sum_{r=1}^{N_s-1}  (N_s-r)    \cos\left(rB_2L_s\xi\right) \right) d\xi 
    \end{aligned}
    \label{eq:K_XS_identical_before_sum}
\end{equation}
The finite sum satisfies:
\begin{equation}
    N_s + 2\sum_{r=1}^{N_s-1}(N_s-r)\cos(r\theta)  =
    \left[\frac{\sin\left(N_s\theta/2\right)}{\sin\left(\theta/2\right)}\right]^2 
    \label{eq:identical_span_sum_identity}
\end{equation}
Therefore:
\begin{equation}
    \begin{aligned}
    G_{\mathrm{NLI,rect}}(f) = &A_{\mathrm{rect}}(f)     \int_{a'c'}^{b'd'}S_{\mathbf p}(B_2\xi)     W_{\mathrm{rect}}(\xi) \\   
    &     \left\{    \left[     \frac{     \sin\left(N_sB_2L_s\xi/2\right)}{\sin\left(B_2L_s\xi/2\right)  } \right]^2  \right\}    d\xi 
    \end{aligned}
    \label{eq:K_XS_identical_closed_sum}
\end{equation}
This expression incorporates all SS and XS contributions into a single integral. When \(B_2=0\), this limit applies throughout the integration domain, corresponding to fully coherent NLI accumulation across all spans.

\subsection{\textbf{Method 2: f-integral first}}
\label{sect:XS_method2}
The frequency integral in $K^{(m,n)}_{\mathrm{NLI,rect}}(f)$ can be evaluated before the two space integral, following the same procedure in Sect.~\ref{sect:SS_method2}:
\begin{equation}
    \begin{aligned}
        K&^{(m,n)}_{\mathrm{NLI,rect}}(f) = \int_{0}^{L_s^{(m)}} \int_{0}^{L_s^{(n)}} \mathbf{p}_{\mathrm{rect}}^{(m)}(z_m) \mathbf{p}_{\mathrm{rect}}^{(n)}(z_n)\\
        &F_{\mathrm{rect}} \!\left[ 4\pi^2(B_m-B_n) + B_2^{(m)} z_m - B_2^{(n)} z_n \right] \,dz_n\,dz_m
    \end{aligned}
    \label{eq:method2_freq_first}
\end{equation}
where $F_{\mathrm{rect}}(\cdot)$ is defined in Eq.~(\ref{eq:FB_definition}) and evaluated analytically. Compared with the corresponding SS expression in Eq.~(\ref{eq:integral_method2}), the argument of $F_{\mathrm{rect}}$ contains the additional accumulated-dispersion difference $4\pi^2(B_m-B_n)$, which accounts for the phase accumulated between the two spans. It remains valid when either local dispersion parameter is zero.

Using the dispersion-domain coordinates introduced in Eq.~(\ref{eq:dispersion_domain_NSP}), we obtain:
\begin{equation}
    \begin{aligned}
    4\pi^2(B_m-B_n)
    +B_2^{(m)}z_m-B_2^{(n)}z_n\\
    =
    4\pi^2(B_m-B_n)+\Delta_u 
    \end{aligned}
    \label{eq:method2_sdeltau}
\end{equation}
Therefore, after integrating over \(\Sigma_u\), Eq.~(\ref{eq:method2_freq_first}) reduces to
\begin{equation}
    \begin{aligned}
    K_{\mathrm{NLI,rect}}^{(m,n)}(f) &=  \int_{\Delta_{u,\text{min}}}^{\Delta_{u,\text{max}}}
        R_{\widetilde{\mathbf p}}^{(m,n)}(\Delta_u) \\
        &F_{\mathrm{rect}}
        \left[
        4\pi^2(B_m-B_n)+\Delta_u
        \right]
        d\Delta_u
    \end{aligned}
    \label{eq:method2_}
\end{equation}
where $\Delta_{u,\min}=u_m^- - u_n^+, \Delta_{u,\max}=u_m^+ - u_n^-$.

\section{Validation}
\label{sect:validation}
Validation is an essential aspect of any modeling effort. Since our goal is to provide a fast and accurate semi-analytical GN-model evaluation framework, our benchmark is the full numerical integration of the GN-model itself, Eq.~(\ref{eq:GNLI_GN}).

To perform the validation, we consider quite different systems, of increasing complexity. They are listed in the following:
%, such as shown in Fig.~\ref{fig:coherent_NLI}, either over a single span or extended to 10 spans, with either pure loss or Raman amplification and ISRS. We assume different signal configurations, from single channel to a massive multi-band configuration.
%:  the first two have identical 100-km spans, while the third has heterogeneous spans, as in \cite{2025_JLT_Jiang} : 
\begin{itemize}
    \item \textbf{Config 1}: single channel at 140~GBaud. The SPP is affected solely by the fiber attenuation, as shown by the blue dashed curve (case 1) in Fig.~\ref{fig:Rp_2case}(a). The link consists of 10 identical spans of 100 km length.
    \item \textbf{Config 2}: five channels with different symbol rates and channel spacings, occupying approximately 500~GHz. The input PSD is shown in Fig.~\ref{fig:config2} (thin solid line), with uniform flat-top values. Channels 2 and 4 have multiple subcarriers. The SPPs of each channel or subcarrier are affected  by fiber attenuation, ISRS and significant backward Raman amplification, similar to the blue solid curve (case 2) in Fig.~\ref{fig:Rp_2case}(a). The WDM signal is propagated over 10 identical spans, of 100 km length.
    \item \textbf{Config 3}: a 150-channel C+L+S system, which comprises 50 channels in each band, all operating at 100~GBaud with a roll-off factor of 0.1 and a channel spacing of 118.75~GHz. The link consists of 10 spans, with span-dependent fiber parameters. The individual channel SPPs account for ISRS and amplification due to three backward Raman pumps. Both the per-channel launch powers and the backward Raman pump powers were jointly optimized to maximize the system throughput. The optimized launch-power profile, shown by the solid black curve in Fig.~\ref{fig:heterogeneous}, is strongly nonuniform across the channels. More details can be found in \cite{2025_JLT_Jiang}.
\end{itemize}

We first focus on a single span in \textbf{Config 1} and \textbf{Config 2}, then we look at the full 10-span links in all three configurations.

\subsection{Config 1 and Config 2, single-span}
\label{sect:test_NLI_SS}
We begin with the signal input of \textbf{Config 1}. The performance metric is the relative error in the NLI power predicted by the cGN framework $P_{\mathrm{NLI}}^\mathrm{\,cGN}$ with respect to the numerically integrated GN-model $P_{\mathrm{NLI}}^{\mathrm{GN}}$, defined as:
\begin{equation}
    \begin{aligned}
       \Delta{P_{\mathrm{NLI}}} = P_{\mathrm{NLI}}^{\mathrm{GN}} - P_{\mathrm{NLI}}^\mathrm{\,cGN}
    \end{aligned}
    \label{eq:performance_indicator}
\end{equation}
All quantities are expressed in decibels. A positive value of $\Delta P_{\mathrm{NLI}}$ indicates that the cGN framework underestimates the GN-model benchmark. 

To further demonstrate the capability of the proposed cGN framework, we compare its performance with that of our previously proposed PCFM1 in ~\cite{2026_JLT_Poggiolini}, using $N_p=5$. PCFM1 estimates the NLI only at the center frequency of the CUT and employs a machine-learning correction factor (MLCF) to improve the accuracy of its underlying closed-form approximation. Moreover, PCFM1 accounts only for SCI and XCI contributions, while MCI contributions are neglected. It is therefore primarily suited to high-dispersion, high-baud-rate systems. Further details regarding PCFM1 and its range of applicability are provided in~\cite{2026_JLT_Poggiolini}.

\begin{figure}
    \centering
    \includegraphics[width=0.9\columnwidth]{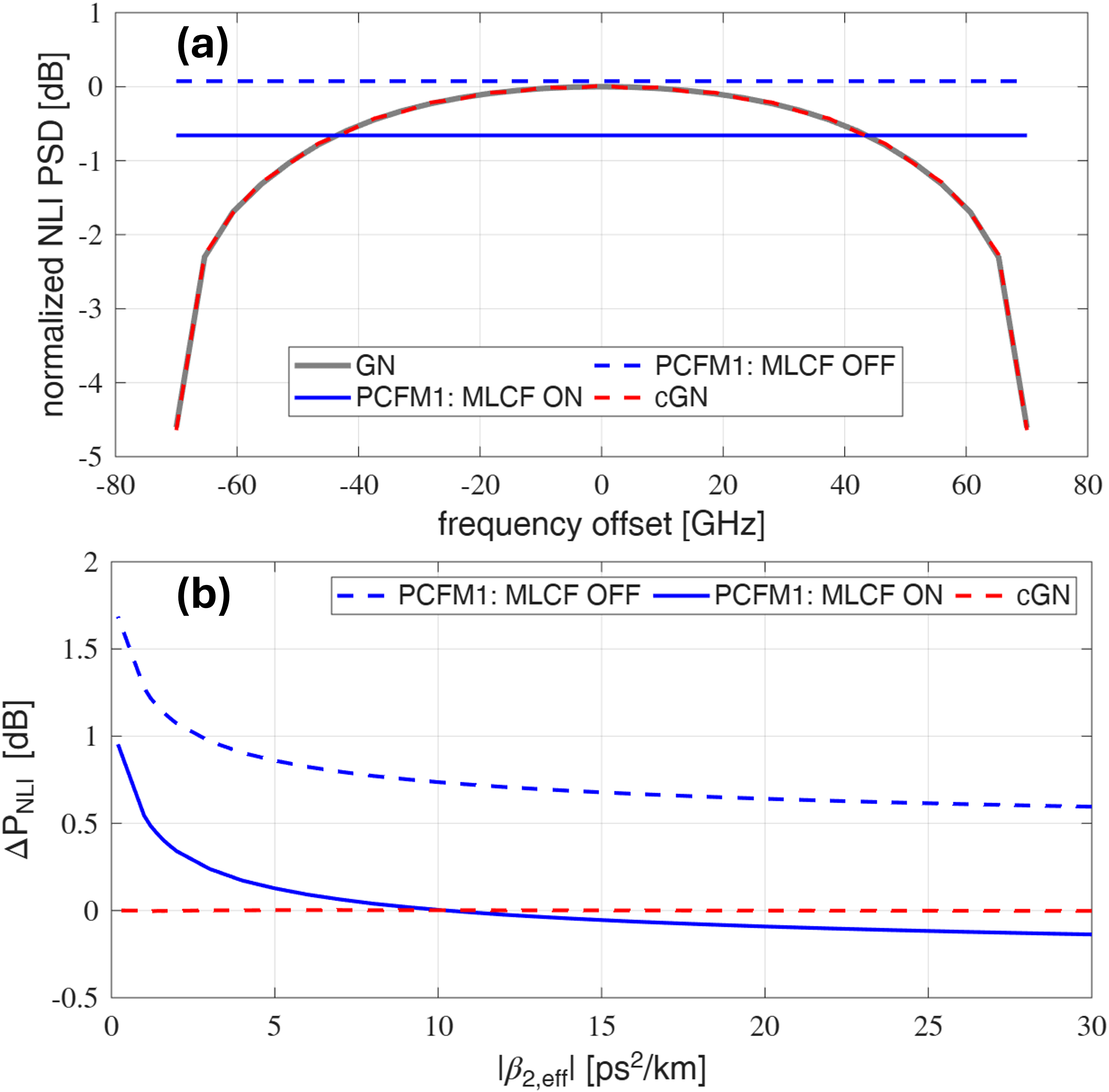}
    \caption{(a) Normalized NLI PSD in a single 140~GBaud channel after 100~km. 
    (b) NLI power error for PCFM1 and cGN, with respect to the GN model over $|\beta_{2,\mathrm{eff}}|\in[0,30]~\mathrm{ps}^2/\mathrm{km}$, in \textbf{Config 1}.}
    \label{fig:test1_SCI_PSD}
\end{figure}

 Figure~\ref{fig:test1_SCI_PSD}(a) compares the NLI PSDs. All curves are normalized to the GN-model value, shown by the gray curve, at the CUT center frequency. Because PCFM1 provides an estimation only at the center frequency, its prediction is represented as a horizontal line across the CUT. The MLCF applies a constant correction to this estimate and therefore does not alter its flat spectral shape. By contrast, the proposed cGN framework directly evaluates the frequency-resolved NLI PSD and accurately reproduces its variation across the CUT.
Fig.~\ref{fig:test1_SCI_PSD}(b) reports $\Delta{P_{\mathrm{NLI}}}$ in Eq.~(\ref{eq:performance_indicator}), over the range $|\beta_{2,\mathrm{eff}}|\in[0,30]~\mathrm{ps}^2/\mathrm{km}$. PCFM1 (dashed blue line) shows a large error vs. the GN-model benchmark, on the order of 1dB. When the MLCF is used, such error is mitigated (solid blue line). However, for low dispersion the MLCF is ineffective and the error goes back up steeply. cGN is instead very accurate across all dispersions, including arbitrarily close to zero and even zero. Such accuracy is not obtained through the complex training of a MLCF, but thanks to the intrinsically accurate formula.

\begin{figure}
    \centering
    \includegraphics[width=1\columnwidth]{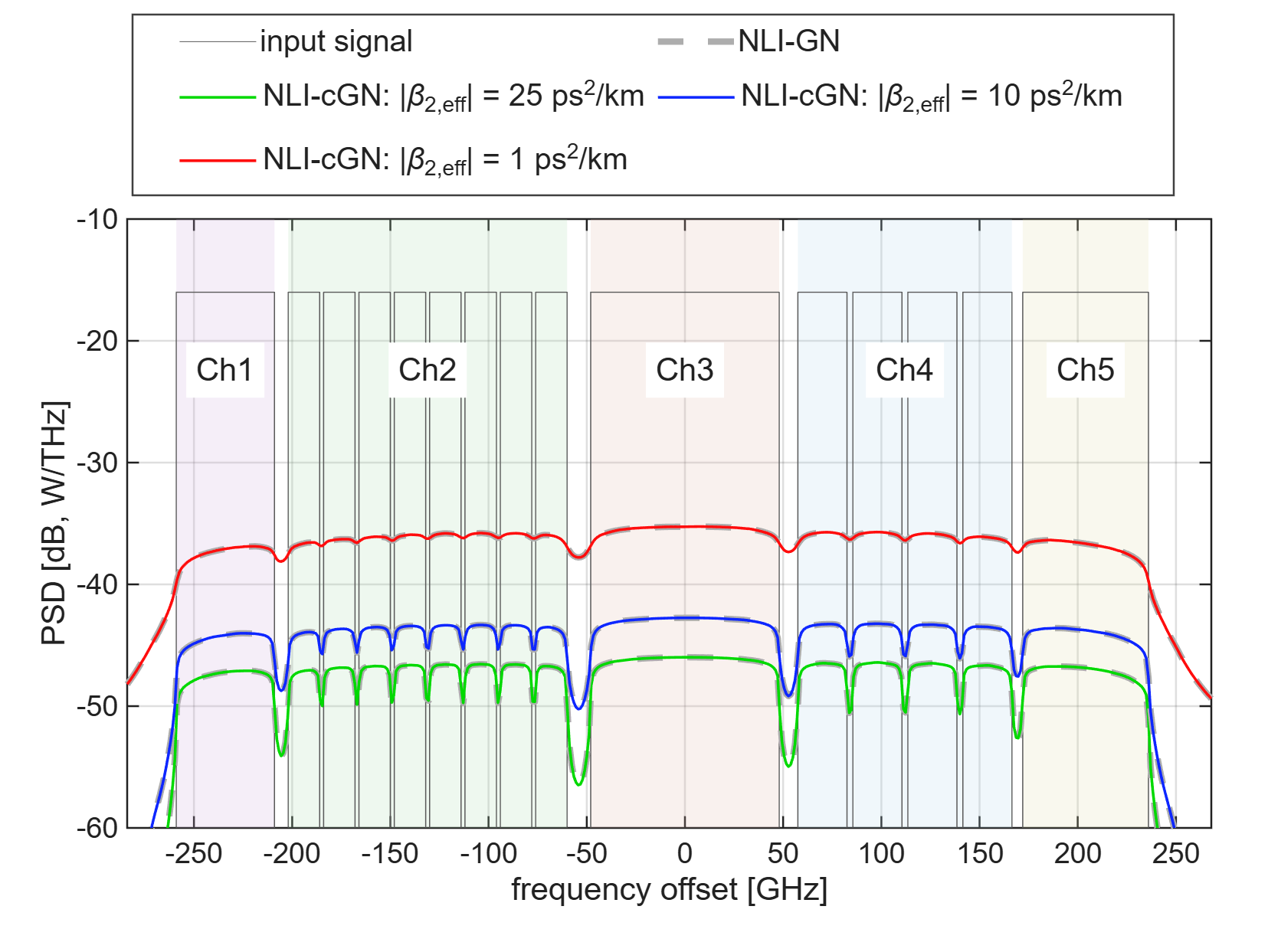}
    \caption{WDM PSD input (black) with different symbol rates and channel spacings, and NLI PSD predicted by the GN-model benchmark (gray curves) and the proposed cGN framework (colored curves), at three different dispersion values, in \textbf{Config 2}.}
    \label{fig:config2}
\end{figure}

We then considered the signal input \textbf{Config 2}. Fig.~\ref{fig:config2} shows the NLI PSD predicted by the GN-model benchmark (gray curves) and the proposed cGN framework (colored curves), at three different dispersion values. The cGN framework accurately captures the NLI in all cases.

\subsection{Multi-span validation}
\label{sect:test_NLI_XS}
We next investigate the coherent interaction between the NLI fields generated in different spans.
\subsubsection{Homogeneous links}
 Ten identical spans are considered, resulting in ten distinct contributions corresponding to different span separations. For each separation $r$, where $r = 0,\ldots,9$, there are $10-r$ terms. 
\begin{figure}
    \centering
    \includegraphics[width=1\columnwidth]{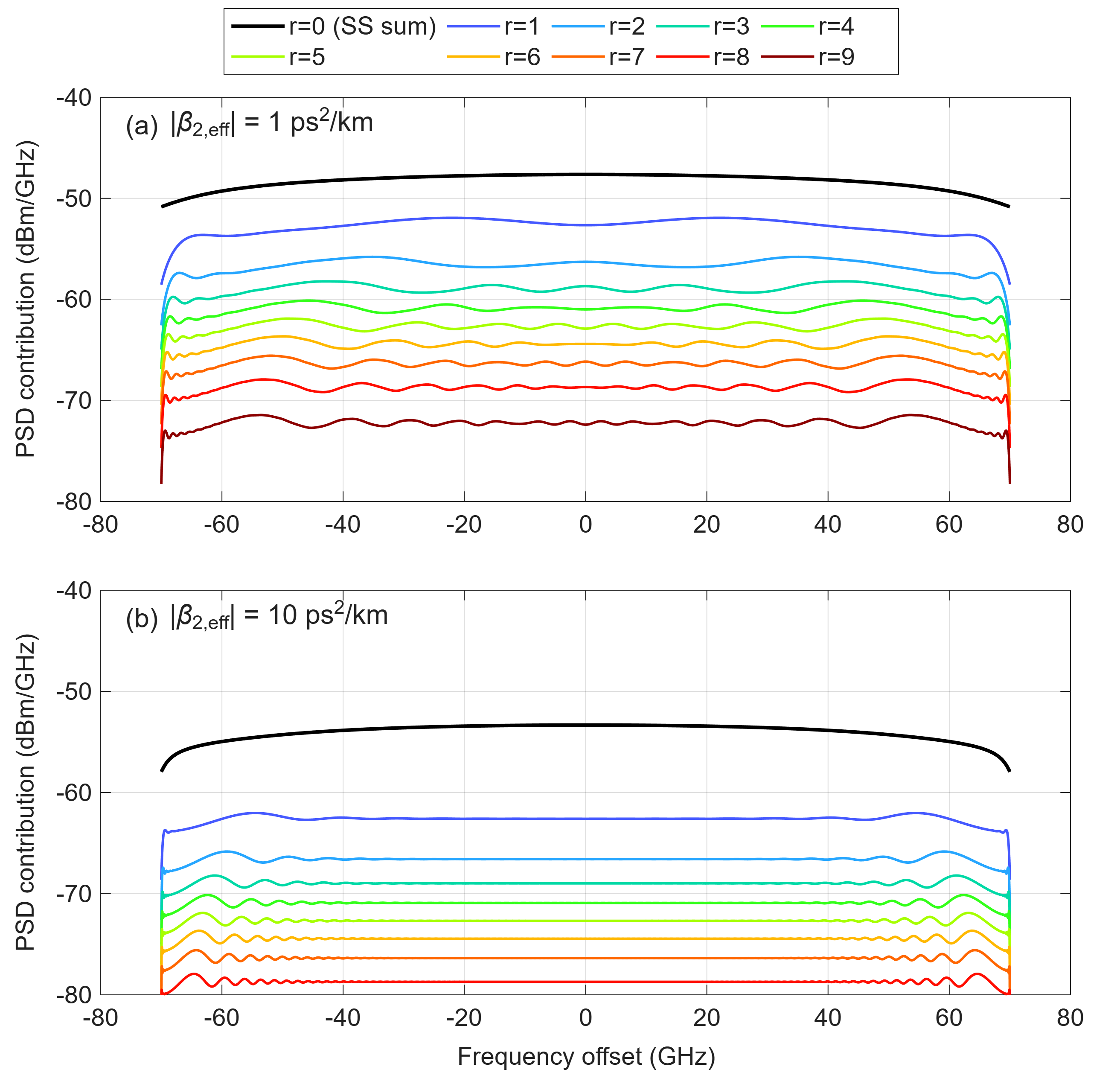}
    \caption{Cross-span SCI PSD contributions for ten identical spans in \textbf{Config 1}: (a) $|\beta_{2,\mathrm{eff}}|=1~\mathrm{ps^2/km}$ and (b) \(|\beta_{2,\mathrm{eff}}|=10~\mathrm{ps^2/km}\). Here, \(r\) denotes the separation between the interacting spans.}
    \label{fig:cross_ten_spans_NLI_PSD}
\end{figure} 

We first focus on \textbf{Config~1}. Fig.~\ref{fig:cross_ten_spans_NLI_PSD} shows all contributions (SS+XS) for two dispersion values. The SS contribution is dominant, while the neighboring-span contribution $r$ provides the most significant coherent correction. The remaining contributions generally decrease as the span separation $r$ increases. Increasing the magnitude of dispersion significantly reduces the XS contributions because the more rapid phase variation leads to weaker coherence between the nonlinear fields generated in different spans.

\begin{figure}
    \centering
    \includegraphics[width=0.9\columnwidth]{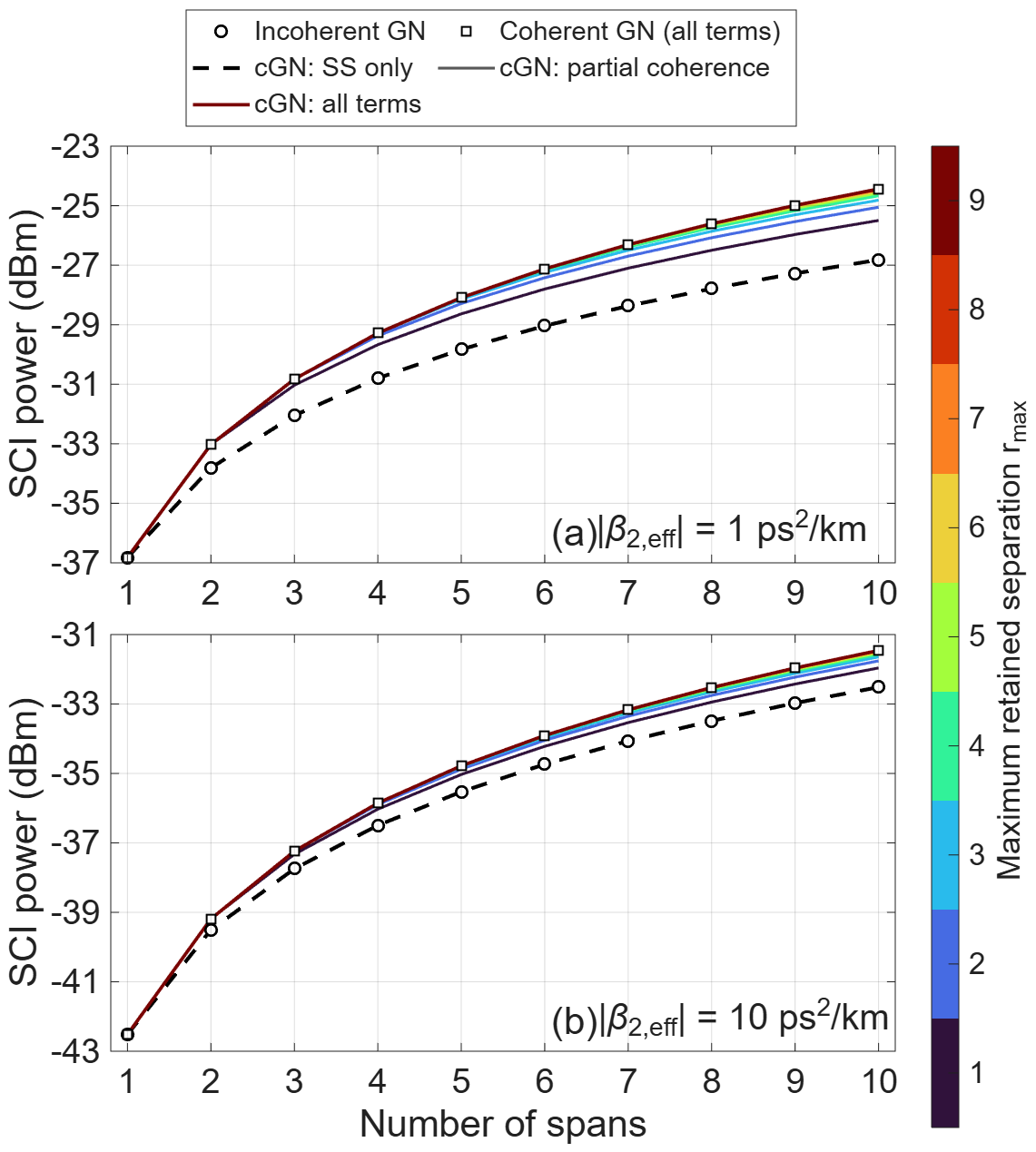}
    \caption{Accumulated SCI power versus the number of spans for the attenuation-only link with (a) \(|\beta_{2,\mathrm{eff}}|=1~\mathrm{ps^2/km}\) and (b) \(|\beta_{2,\mathrm{eff}}|=10~\mathrm{ps^2/km}\). Circle and square markers denote the incoherent and coherent GN results, respectively. The black dashed curve shows the cGN result retaining only the SS terms, while the colored solid curves add XSterms up to the maximum span separation $r_{\max}$, as indicated by the color bar, in \textbf{Config 1}.}
    \label{fig:SCI_power_Config1}
\end{figure} 

Figure~\ref{fig:SCI_power_Config1} compares the accumulated SCI power predicted by the proposed cGN framework and the GN-model benchmark. The power increases incoherently when only the SS terms are considered, as shown by the dashed line.  As the XS contributions for $r=1,\ldots,9$ are progressively included, the cGN prediction approaches and eventually overlaps with that of the coherent GN model. After ten spans, the actual SCI power exceeds the incoherent result by approximately 2.40 dB for $\beta_2=-1~\mathrm{ps^2/km}$, and 1.07 dB for $\beta_2=-10~\mathrm{ps^2/km}$. This confirms that larger accumulated dispersion reduces XS coherence. 
Notably, retaining only the SS terms and the nearest-neighbor XS contribution $r=1$, the purple curve, already captures about half of the total coherent enhancement: 1.32 dB vs. 2.40 dB for $\beta_2=-1~\mathrm{ps^2/km}$, and 0.55 dB vs. 1.07 dB for $\beta_2=-10~\mathrm{ps^2/km}$. From 3 to 10 spans, this curve remains approximately parallel to the incoherent GN curve.

\begin{figure}
    \centering
    \includegraphics[width=0.95\columnwidth]{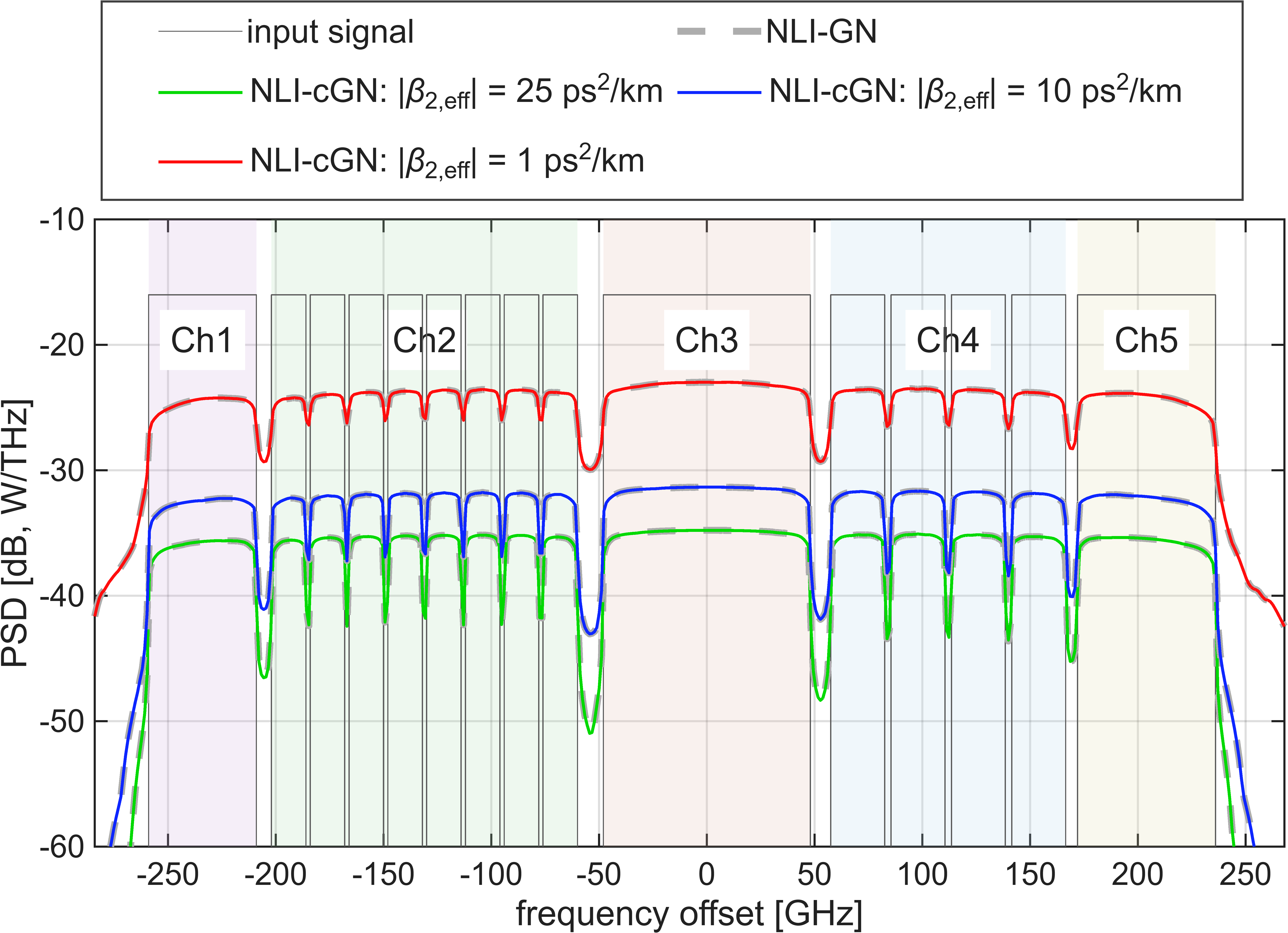}
    \caption{WDM PSD input (black), together with the total 10-span NLI PSDs predicted by the GN-model benchmark (gray curves) and the proposed cGN framework (colored curves) for three effective-dispersion values, in \textbf{Config~2}.}
    \label{fig:config2_10spans}
\end{figure}

We then move to \textbf{Config~2}. Fig.~\ref{fig:config2_10spans} compares the 10-span NLI PSDs predicted by the GN-model benchmark and the cGN framework for three different dispersion values. The cGN curves closely overlap the corresponding GN model results over the entire frequency range. For all cases, the maximum discrepancy between the cGN and GN predictions remains below $0.065$~dB, while the difference in NLI power does not exceed $0.004$~dB. These results confirm that the proposed cGN framework accurately predicts the coherently accumulated NLI including the low-dispersion regime.

\subsubsection{Heterogeneous Links}
We use \textbf{Config~3}, adapted from Fig.~4(a) of \cite{2025_JLT_Jiang}, to assess the proposed formulation under span-dependent transmission conditions.

\begin{figure}
    \centering
    \includegraphics[width=1\columnwidth]{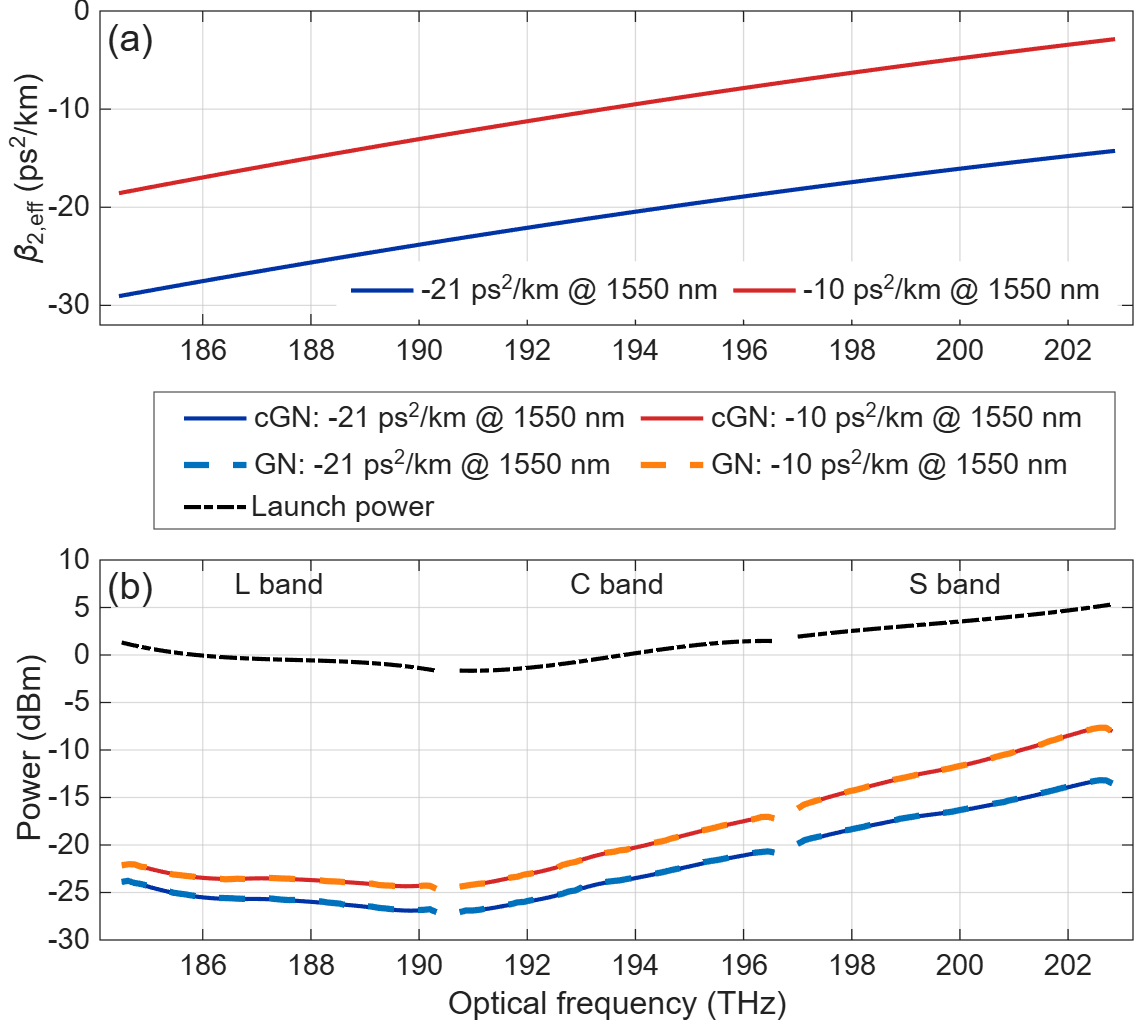}
    \caption{(a) Span-averaged dispersion curves for two configurations with $\beta_{2,\mathrm{eff}}$ of approximately -21 and -10 $\mathrm{ps^2/km}$ at 1550 nm. (b) Per-channel launch power (black dash-dotted line) and NLI power predicted by cGN (solid lines) and numerical GN integration (dashed lines) for the heterogeneous 150-channel C+L+S link under the two dispersion configurations, in \textbf{Config~3} .}
    \label{fig:heterogeneous}
\end{figure}
The ten-span link is constructed from five experimentally characterized spans that are repeated twice. The span-dependent length, wavelength-dependent attenuation, dispersion parameters, nonlinear coefficient, and channel-resolved SPPs are retained separately. Fig.~\ref{fig:heterogeneous} compares the cGN predictions with the GN-model benchmark for all 150 channels of the heterogeneous C+L+S system under two dispersion settings: the original profile in \cite{2025_JLT_Jiang}, with a span-averaged $\beta_{2,\mathrm{eff}}(1550~\mathrm{nm})$ of approximately $-21~\mathrm{ps^2/km}$, and a shifted profile with $\beta_{2,\mathrm{eff}}(1550~\mathrm{nm})=-10~\mathrm{ps^2/km}$. The launch powers and all other link parameters are unchanged. The mean and maximum absolute differences between the cGN framework and the GN-model benchmark are 0.062 and 0.098~dB for the original dispersion profile, and 0.025 and 0.071~dB for the shifted profile, respectively. Reducing the dispersion magnitude increases the predicted NLI across all channels. The good agreement is maintained despite the span-dependent fiber parameters and strongly frequency-dependent Raman power evolution, demonstrating the applicability of the proposed formulation to heterogeneous multi-span links. 

Decomposition of the cGN predictions into SCI, XCI, and MCI shows that SCI and XCI dominate the NLI in both cases. The MCI fraction increases from approximately 0.17\% to 1.10\% as the dispersion at 1550~nm changes from $-21$ to $-10~\mathrm{ps^2/km}$. For the shifted profile, the per-channel MCI fraction remains below 0.16\% across the L and C bands but reaches approximately 3.30\% at the high-frequency edge of the S band. Neglecting MCI would underestimate the total NLI by at most 0.146~dB. These results show that the contribution of MCI increases as the magnitude of dispersion decreases.

\subsection{Numerical convergence and computational efficiency}

Having established the agreement between cGN and the GN-model benchmark for both homogeneous and heterogeneous links, we now examine the numerical accuracy and computational cost. The original two-dimensional frequency integral in each island is reduced to a sum of one-dimensional integrals over a rectangular partition in cGN. For a fixed partition, the correlation-domain transformations do not introduce additional analytical approximation. We therefore vary the number of rectangles per island, $N_{\mathrm{rect}}$, while keeping all other settings fixed, to examine the accuracy and computational cost.

\begin{figure}
    \centering
    \includegraphics[width=0.95\columnwidth]{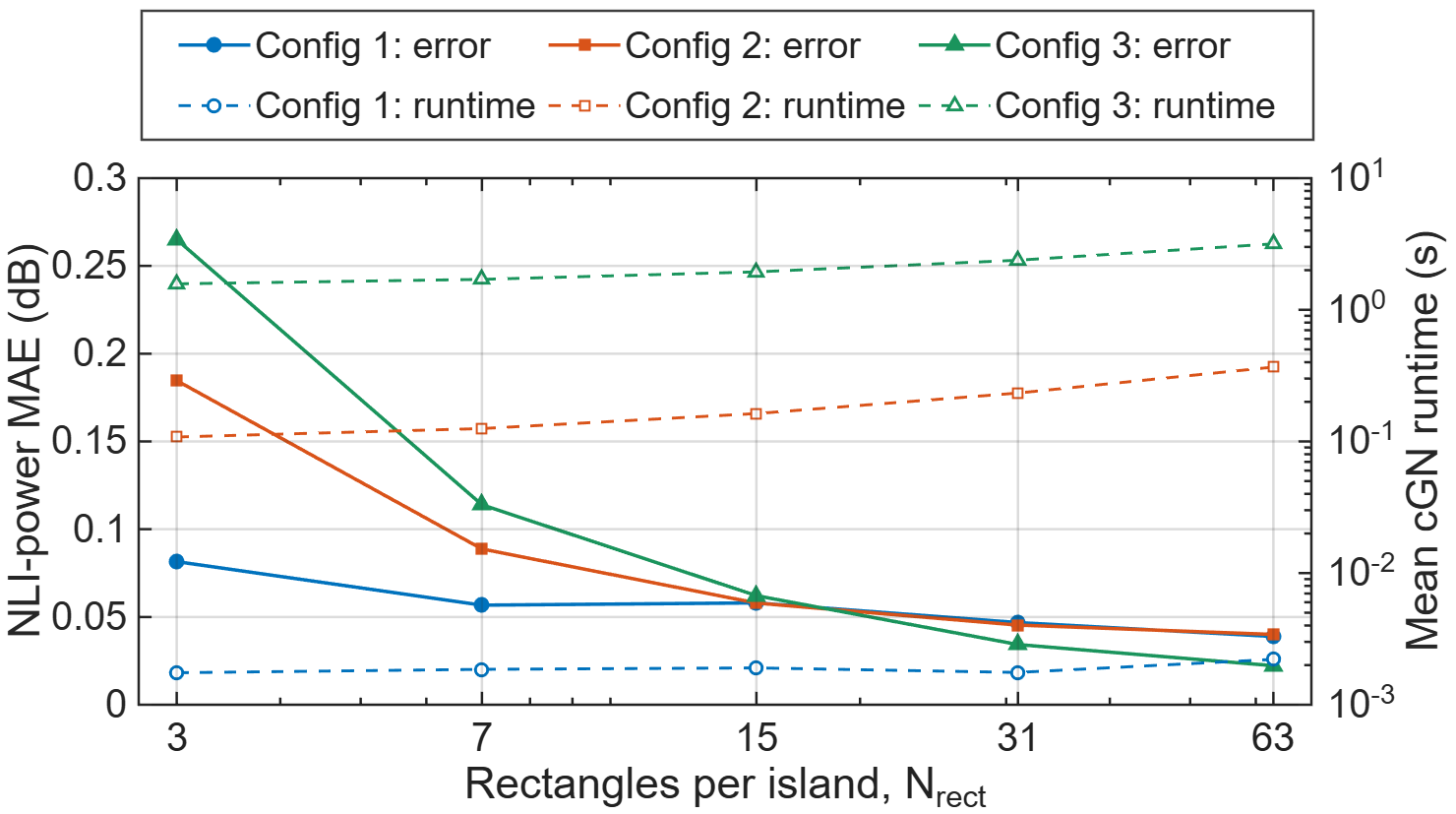}
    \caption{Accuracy--cost trade-off as a function of the number of rectangles per island for the multi-span links in \textbf{Config~1--3}, Solid curves show MAE in the NLI power compared to the GN-model benchmark, and dashed curves show the cGN runtime.}
    \label{fig:convergence_runtime}
\end{figure}

Figure~\ref{fig:convergence_runtime} illustrates this trade-off for the multi-span links considered in {Config~1-3}. The kernel resolution and output-frequency quadrature are kept fixed in this comparison. The accuracy is quantified by the mean absolute error (MAE) in NLI power compared to the GN-model benchmark. The runtime refers to the calculation under the same computational environment.

As $N_{\mathrm{rect}}$ increases, MAE decreases systematically in all three configurations, while the runtime generally increases. {Config~1} and {Config~2} converge relatively rapidly toward their respective GN-model benchmarks. {Config~3} exhibits slower convergence because it involves a substantially more demanding wide-band calculation over a heterogeneous multi-span link. Nevertheless, The consistent reduction in MAE demonstrates that the accuracy can be systematically controlled by refining the partition.

The marginal accuracy improvement becomes progressively smaller at large $N_{\mathrm{rect}}$, particularly for the simpler configurations. The rectangle count can therefore be selected according to the requirements of the application: a relatively coarse partition is suitable for rapid system-level evaluations, whereas a finer partition can be used when higher numerical accuracy is required. For {Config~3}, using 17 frequency points and 15 rectangles per island, cGN evaluated the NLI across all 150 channels in 1.9~s on an Ryzen 7 8845HS processor using 16 threads. The corresponding GN-model benchmark required 4689.2~s, yielding a speedup greater than $2000\times$ while maintaining an MAE of 0.062~dB, as summarized in Table~\ref{tab:runtime}.

The two evaluation methods are mathematically equivalent but lead to different numerical implementations.  Method 1 was used for the production calculations because its reusable spatial Fourier-transform tables resulted in a lower computational cost. Method 2 provides an independent numerical cross-check and a complementary correlation-domain interpretation. Both formulations remain well-defined in the zero- and near-zero-dispersion limits.

The cGN framework is available from the European Union Zenodo website \cite{2026_zenodo_Jiang}.

\begin{table}[t]
\centering
\caption{Accuracy and runtime in \textbf{Config~3}.}
\begin{tabular}{lccc}
\hline
Method & Numerical setting & Runtime & MAE\\
\hline
 GN-model benchmark &
17 frequency points &
4689.2 s & Reference\\
cGN framework &
17 points, 15 rectangles &
1.9 s &
0.062 dB\\
\hline
\label{tab:runtime}
\end{tabular}
\end{table}

\section{Conclusion}
We introduced a semi-analytical cGN framework for the efficient evaluation of NLI in general multi-span optical systems. The framework supports span-dependent fiber parameters, arbitrary channel arrangements and symbol rates, multi-subcarrier signals, and general SPPs arising from fiber attenuation, ISRS, distributed Raman amplification, and lumped gain or loss. By partitioning the GN-model integration domain into sufficiently small rectangles, the contribution of each rectangle is decomposed into SS and XS terms. The former are governed by the spatial autocorrelation, or equivalently the energy spectrum, of the nonlinear-stimulus profile, whereas the latter are described by cross-correlations and cross-spectra of the profiles associated with two different spans.

Two equivalent semi-analytical evaluation methods were derived. Performing the space integrations first reduces each rectangular contribution to a one-dimensional integral over the frequency-product variable. Alternatively, performing the frequency-domain integrations first yields a one-dimensional dispersion-lag integral involving the rectangular coherence-response function. 

The proposed cGN framework was validated against the GN-model benchmark for single- and multi-channel systems, including nonuniform channel arrangements, multi-subcarrier signals, complex SPPs, and low- and conventional-dispersion regimes. The cGN framework accurately reproduced both the NLI spectra and their coherent accumulation over multiple spans. In the more demanding heterogeneous 150-channel C+L+S system, the maximum absolute difference remained below $0.1~\mathrm{dB}$, reaching $0.098~\mathrm{dB}$, while the mean absolute difference was $0.062~\mathrm{dB}$. Under the same computational conditions, the cGN evaluation required $1.9~\mathrm{s}$, compared with $4689.2~\mathrm{s}$ for refined numerical GN integration, corresponding to a speedup of more than $2000\times$. 

These results confirm that the proposed formulation retains the accuracy of the GN-model benchmark while providing an efficient and physically transparent approach for NLI evaluation in general multi-band optical systems.

%\clearpage

\end{document}